\documentclass[english,aps,preprint, superscriptaddress]{revtex4}
\usepackage[T1]{fontenc}
\usepackage[latin9]{inputenc}
\usepackage{mathrsfs}
\usepackage{amsmath}
\usepackage{amssymb}
\usepackage{cancel}

\makeatletter
\@ifundefined{textcolor}{}
{%
 \definecolor{BLACK}{gray}{0}
 \definecolor{WHITE}{gray}{1}
 \definecolor{RED}{rgb}{1,0,0}
 \definecolor{GREEN}{rgb}{0,1,0}
 \definecolor{BLUE}{rgb}{0,0,1}
 \definecolor{CYAN}{cmyk}{1,0,0,0}
 \definecolor{MAGENTA}{cmyk}{0,1,0,0}
 \definecolor{YELLOW}{cmyk}{0,0,1,0}
}

\usepackage{slashed}
\usepackage{babel}

\makeatother

\usepackage{babel}
\begin{document}
\title{$\mathcal{N}=1$ $\mathcal{D}=3$ Lifshitz-Wess-Zumino model: A paradigm of reconciliation between Lifshitz-like operators and supersymmetry}
\author{E. A. Gallegos} 
\email{egallegoscollado@gmail.com}
\affiliation{Instituto de Ciência e Tecnologia, Universidade Federal de Alfenas,Campus Avançado de Poços de Caldas, 37701-970, MG, Brazil}
\begin{abstract}
By imposing the weighted renormalization condition and the (super)symmetry
requirements, we construct a Lifshitz-like extension of the three-dimensional
Wess-Zumino model, with dynamical critical exponent $z=2$. In this
context, the auxiliary field $F$ plays a key role by introducing
the appropriate Lifshitz operator in the bosonic sector of the theory,
avoiding so undesirable time-space mixing derivatives and inconsistencies
concerning the critical $z$ exponent, as reported in the literature.
The consistency of the proposed model is verified by building explicitly
the susy algebra through the Noether method in the canonical formalism.
This component-field Lifshitz-Wess-Zumino model is in addition rephrased
in the Lifshitz superspace, a natural modification of the conventional
one. The one-loop effective potential is computed to study the possibility
of symmetry breaking. It is found that supersymmetry remains intact
at one-loop order, while the $U(1)$ phase symmetry suffers a spontaneous
breakdown above the critical value of the renormalization point. By
renormalizing the one-loop effective potential within the cutoff regularization
scheme, it is observed an improvement of the UV behavior of the theory
compared with the relativistic Wess-Zumino model.
\end{abstract}
\maketitle

\section{INTRODUCTION\label{sec:Sec1}}

In the last few years, there has been an increasing interest in the
study of quantum field theories with higher spatial derivative terms,
known in the literature as Lifshitz-type quantum field theories. One
of the reasons of this interest is the improved UV behavior of the
propagators at high energies, without spoiling the unitarity of the
theory, due to the introduction of higher spatial derivatives (Lifshitz-like
operators) in the kinetic part of the Lagrangian. The renormalizability
as well as the unitarity of this kind of theory are ensured by the
so-called weighted renormalization condition \cite{Anselmi =000026Halat (2007),Anselmi (2009)}.
This condition in turn requires an anisotropy (see Eq. \ref{eq:II.3})
between space and time coordinates so that the Lorentz symmetry is
lost in the UV region. It is believed, however, that this symmetry
should emerge at low energies. This subtle issue was investigated
in several papers, see for example \cite{Iengo et al (2009),Gomes et al (2015)}.

The simplest Lifshitz scalar theory with critical exponent $z=2$
was proposed long ago with the intention of explaining the second-order
phase transitions in condensed-matter systems \cite{Lifshitz (1941)}.
Since then several generalizations of this prototype have appeared
in condensed-matter physics, high energy physics and gravity (see
\cite{Alexandre (2011)} and references therein). In this last context,
the Ho$\check{\text{r}}$ava-Lifshitz gravity proposed by Ho$\check{\text{r}}$ava
\cite{Horava (2009)} is arguably one of the most important reincarnations
of the Lifshitz's ideas. The Ho$\check{\text{r}}$ava's proposal is
simply a quantum field theory of gravity of the Lifshitz type with
critical exponent $z=3$ which violates the Lorentz invariance, due
to the introduction of Lifshitz-like operators, in favor of its renormalizability
at high energies. This theory of course will make physical sense if
the restoration of the Lorentz symmetry occurs in the IR region. At
the present time this is still an open subject of investigation \cite{Wang (2017)}.
Regarding this point, an interesting variant of the HL-gravity was
proposed and studied in \cite{Cognola et.al. (2016)}. In this gravity's
proposal, the diffeomorphism invariance is broken dynamically in the
UV region in order to avoid some instability problems inherent in
HL-gravity.

On the other hand, the implementation of Lifshitz-like operators in
supersymmetric field theories is not a trivial task and so far there
does not exist a natural method of doing it. This problem was faced
in \cite{Redigolo (2012),Gomes et al (2014),Gomes.Quiruga (2015)}
by employing the superfield formalism in four spacetime dimensions.
Nevertheless, some inconsistencies concerning the ill-definedness
of the critical exponent $z$ and the appearance of undesirable time-space
mixing derivatives were observed in the Lifshitz-like constructions
proposed in \cite{Gomes et al (2014)}. 

In order to see more clearly what is happening and eventually uncover
the real roots of these inconsistencies, we tackle the problem of
constructing a Lifshitz-like extension of the Wess-Zumino model in
the three dimensional component formalism, i.e., without employing
the conventional superfield formalism. The three-dimensional framework
constitutes an excellent theoretical laboratory for an in-depth study
of these four-dimensional setbacks, since the notion of chirality
does not exist in odd dimensions and so three-dimensional susy theories
become simpler, conserving, however, the main features of their four-dimensional
counterparts. To complete our research, in Appendix \ref{sec:Appendix.B},
we construct the four dimensional Lifshitz-Wess-Zumino model with
critical exponent $z=2$. Here we show how the inconsistencies mentioned
above can be avoided by modifying adequately the usual superfield
formalism (as suggested from our three-dimensional study) and by using
the weighted renormalization condition as a guide.

In the body of this paper, we show in detail that the insertion of
Lifshitz-like operators according to the weighted renormalization
condition is completely compatible with supersymmetry and the well-definedness
of the critical exponent $z$. Furthermore, in the superspace reformulation
of the proposed model (the Lifshitz-Wess-Zumino model), we show that
the conventional superfield formalism is inappropriate for the formulation
of supersymmetric theories of the Lifshitz type. It is not hard to
notice that the conventional superfield formalism invented by Salam
and Strathdee \cite{Salam =000026 Strathdee} for constructing relativistic
susy theories does violate the weighted power-counting criterion.
Indeed, since the susy-relativistic covariant derivative, 
\begin{equation}
D_{\alpha}=\partial_{\alpha}+i\left(\gamma^{\mu}\right)_{\alpha\beta}\theta^{\beta}\partial_{\mu}=\partial_{\alpha}+i\left(\gamma^{0}\right)_{\alpha\beta}\theta^{\beta}\partial_{0}+i\left(\gamma^{i}\right)_{\alpha\beta}\theta^{\beta}\partial_{i},
\end{equation}
embodies the time and space derivatives with the same weight, this
susy covariant object does not obey the anisotropic scaling rules
(\ref{eq:II.3}), with $z>1$. This fact is non-negotiable and illustrates
the necessity of modifying the conventional superfield formalism before
employing it in the construction of Lifshitz-like susy theories. This
minor and necessary modification was carried out in Eq. (\ref{eq:II.22b})
in order to express the component-field Lifshitz-Wess-Zumino model
in terms of the superfield language (i. e., in the Lifshitz superspace
as we call it). 

In the same spirit as conventional (i.e. without susy) Lifshitz-type
quantum field theories \cite{Anselmi =000026Halat (2007),Anselmi (2009)},
we attempted to split the five-dimensional superspace $\mathcal{SM}_{5}$
into the product of two disjoint submanifolds (supersectors): $\mathcal{SM}_{t}\times\mathcal{SM}_{s}$,
where $\mathcal{SM}_{t}$ stands for the supertime manifold and $\mathcal{SM}_{s}$
the superspatial one. The goal of this separation is to create an
environment more adequate and natural for constructing Lifshitz-type
susy theories in which the time derivative $\partial_{0}$ and the
spatial derivative $\partial_{i}$ could live (or act) in entirely
distinct supersectors. As is shown in Appendix \ref{sec:Appendix.A},
such a complete separation is unworkable without the introduction
of extra Grassmannian coordinates. Needless to say, that this procedure
would require a complete reformulation of the conventional superfield
formalism. This issue perhaps deserves further attention, in particular,
in the construction of gauge susy theories of the Lifshitz type. For
the moment, this is beyond the scope of this paper.

Finally, the effective potential of the Lifshitz-Wess-Zumino model
was computed at one-loop order. The purpose of this calculation is
twofold: to investigate the possibility of susy breaking due to the
Lifshitz-like operators implemented in the conventional Wess-Zumino
model and to understand how the UV improvement occurs in susy theories
of the Lifshitz type. By analyzing the stationary conditions of the
one-loop effective potential, we show that supersymmetry remains intact,
while the $U(1)$ phase symmetry suffers a spontaneous breakdown above
the critical value of the renormalization point. On the other hand,
the improved UV behavior of the theory becomes evident by introducing
a two-dimensional cutoff $\left(\Lambda\right)$ to regularize the
one-loop effective potential. This result, nevertheless, depends on
the exact cancellation of the quadratic divergences between the bosonic
and fermionic contributions (see comments below Eq. (\ref{eq:III.6b})).
In scalar field theories (without susy), the one-loop effective potential
was computed in \cite{Eune et al (2011)}.

Our paper is organized as follows. In Sec. \ref{sec:Sec2} we construct
a Lifshitz-like extension of the three dimensional Wess-Zumino model
with critical exponent $z=2$. This is done by imposing the weighted
renormalization condition and the (super)symmetry requirements. Here
the susy algebra is explicitly constructed by using the Noether method.
Furthermore, our field-component model is rephrased in the Lifshitz
superspace. In Sec. \ref{sec:Sec3}, the one-loop effective potential
is calculated and their mimina analyzed. To investigate the structure
of the UV divergences, the one-loop effective potential is regularized
by using a two-dimensional cutoff $\left(\Lambda\right)$. Sec. \ref{sec:Sec4}
contains our main results. Supplementary informations are available
in Appendices \ref{sec:Appendix.A}, \ref{sec:Appendix.B}, and \ref{sec:Appendix.C}.
In Appendix \ref{sec:Appendix.B}, in particular, we construct the
$z=2$ Lifshitz version of the Wess-Zumino model in four spacetime
dimensions. This is done directly in the Lifshitz superspace by applying
the weighted renormalization condition. The Appendix \ref{sec:Appendix.C},
on the other hand, is very odd since in it we construct the Lifshitz
extension of the susy Maxwell theory in three dimensions, an issue
outside of the scope of this work. However, this is treated here to
reinforce our belief that the current superfield formulation of susy
theories is inadequate for constructing susy theories of Lifshitz
type. 

\section{$\mathcal{N}=1$ $\mathcal{D}=3$ Lifshitz-Wess-Zumino model\label{sec:Sec2}}

In this section, we are going to construct a Lifshitz-like extension
of the Wess-Zumino model in $\left(1+2\right)$ spacetime dimensions,
by imposing the weighted renormalization condition and the (super)symmetry
requirements. In order to do this, we split the three dimensional
spacetime manifold $\mathcal{M}_{3}$ into the product $\mathbb{R}\otimes\mathcal{M}_{2}$,
where $\mathbb{R}$ represents the one-dimensional time submanifold
and $\mathcal{M}_{2}$ the two-dimensional spatial submanifold. Therefore,
in this work, we shall construct an action $S$ which is invariant
under spatial rotations, space and time displacements and supersymmetry.
The rigid phase $U\left(1\right)$ symmetry is also imposed. Note
that the original Lorentz group $SO\left(1,\,2\right)$ after this
separation becomes simply the group of spatial rotations, i.e. the
group $SO\left(2\right)$. 

Our starting point will be the action
\begin{eqnarray}
S & = & \int dtd^{2}x\left[-\left(\partial^{0}\bar{\varphi}\partial_{0}\varphi+a^{2}\partial^{i}\bar{\varphi}\partial_{i}\varphi\right)+i\bar{\psi}\cancel{\partial}_{0}\psi+ia\bar{\psi}\cancel{\partial}_{i}\psi+\bar{F}F+\mathcal{L}_{int}\right],\label{eq:II.1}
\end{eqnarray}
where $\varphi$ is a complex scalar field, $\psi_{\alpha}$ a complex
(Euclidean) spinor field, and $F$ a complex auxiliary field and where
$a$ is a dimensionless parameter whose weight is $\left[a\right]_{w}=z-1$.
From now on we shall adopt the notation of \cite{Gates et al (1983)}.
In particular, $i\bar{\psi}\cancel{\partial}_{\mu}\psi$ means $i\bar{\psi}^{\alpha}\left(\gamma^{\mu}\right)_{\alpha}^{\,\,\,\beta}\partial_{\mu}\psi_{\beta}$.
Note that we maintain the residual Lorentz notation for the time and
space derivatives, namely $\partial^{0}=-\partial_{0}$ and $\partial^{i}=+\partial_{i}$.
As will be seen later, the auxiliary field $F$ plays a key role in
our construction. In fact, in addition to its usual role of making
susy manifest off-shell, $F$ shall allow us to introduce a higher
space derivative in the scalar sector of the theory without altering
its susy algebra.

Switching off the interaction Lagrangian $\mathcal{L}_{int}$, i.e.
taking $\mathcal{L}_{int}=0$ in (\ref{eq:II.1}), it is easy to show
that the resulting free action is invariant under the following susy
transformations
\begin{eqnarray}
\delta\varphi & = & -\epsilon\psi\nonumber \\
\delta\psi & = & \epsilon F-i\epsilon\cancel{\partial}_{0}\varphi-ia\epsilon\cancel{\partial}_{i}\varphi\label{eq:II.2}\\
\delta F & = & -i\epsilon\cancel{\partial}_{0}\psi-ia\epsilon\cancel{\partial}_{i}\psi,\nonumber 
\end{eqnarray}
where $\epsilon$ is a Grassmann $x$-independent parameter.

Before proceeding with the construction of $\mathcal{L}_{int}$, it
is necessary to state clearly the weighted renormalization condition
(wrc) \cite{Anselmi =000026Halat (2007),Anselmi (2009)}. If one writes
$\mathcal{L}_{int}$ of a given theory as $\mathcal{L}_{int}=\sum_{i}g_{i}\mathcal{V}_{i}$,
where $g_{i}$ label the coupling constants and $\mathcal{V}_{i}$
the interaction vertices, this condition simply says that the theory
is renormalizable by weighted power counting iff the weighted scaling
dimension, the weight for short, $\left[g_{i}\right]_{w}$ of each
coupling constant $g_{i}$ is greater or equal to zero, i.e., $\left[g_{i}\right]_{w}\geq0$.
Setting $\left[x^{0}=t\right]_{w}=-z$ and $\left[x^{i}\right]_{w}=-1$,
the weight $\left[\mathcal{O}\right]_{w}$ of any object $\mathcal{O}$
is determined by enforcing the action $S$ to be weightless. In terms
of the vertices, since the Lagrangian $\mathcal{L}$ weighs $d+z$,
where $d$ denotes the spatial dimensions, the wrc asserts that a
vertex $\mathcal{V}_{i}$ is weighted renormalizable iff $\left[\mathcal{V}_{i}\right]_{w}\leq d+z$.
Following the nomenclature adopted in the literature, we call a vertex
weighted marginal when this weighs $\left[\mathcal{V}_{i}\right]_{w}=d+z$,
weighted relevant when $\left[\mathcal{V}_{i}\right]_{w}<d+z$ and
weighted irrelevant when $\left[\mathcal{V}_{i}\right]_{w}>d+z$.

It should be noted that the weighted assignment in Lifshitz field
theories is equivalent to demand the invariance of the action under
the following anisotropic scale transformations,
\begin{equation}
x^{i}\rightarrow\xi x^{i}\qquad\qquad\qquad t\rightarrow\xi^{z}t,\label{eq:II.3}
\end{equation}
where $z$ is the well-known critical exponent which measures the
degree of anisotropy between space and time. Moreover, notice that
the weighted scaling dimension coincides with the usual mass one in
natural units ($\hbar=1=c$) when $z=1$.

To construct $\mathcal{L}_{int}$ we first observe that any vertex
$\mathcal{V}$ in it must have the structure $\mathcal{V}\sim\left(\partial_{i}\right)^{N_{\partial_{i}}}\psi^{N_{\psi}}\bar{\psi}^{N_{\bar{\psi}}}\varphi^{N_{\varphi}}\bar{\varphi}^{N_{\bar{\varphi}}}F^{N_{F}}\bar{F}^{N_{\bar{F}}}$,
where $N_{p}$ represents the number of objects of the $p$ type.
Since the fields in the action (\ref{eq:II.1}) for arbitrary $z$
weigh $\left[\varphi\right]_{w}=\left(2-z\right)/2$, $\left[\psi_{\alpha}\right]_{w}=1$,
and $\left[F\right]_{w}=\left(2+z\right)/2$, one easily sees by imposing
the wrc which $\mathcal{V}$ is renormalizable by weighted power counting
iff the condition
\begin{equation}
N_{\partial_{i}}+\mathcal{N}_{\psi}+\frac{\left(2-z\right)}{2}\mathcal{N}_{\varphi}+\frac{\left(2+z\right)}{2}\mathcal{N}_{F}\leq2+z,\label{eq:II.3.1}
\end{equation}
is satisfied. Here $\mathcal{N}_{X}=N_{X}+N_{\bar{X}}$, with $X=\varphi,\,F,\,\psi$.
This condition along with those which result from imposing the symmetry
requirements restrict strongly the form of $\mathcal{L}_{int}$ .
Note in particular that the spatial rotational $SO\left(2\right)$
symmetry of the Lagrangian $\mathcal{L}$ demands a complete spinor/spatial
index contraction as well as that $\mathcal{N}_{\psi}=N_{\psi}+N_{\bar{\psi}}=\text{even number}$.
The rigid phase $U(1)$ symmetry, on the other hand, implies that
$N_{\psi}+N_{\varphi}+N_{F}=N_{\bar{\psi}}+N_{\bar{\varphi}}+N_{\bar{F}}$.
It should be noted that whether the polynomiality requirement of the
Lagrangian $\mathcal{L}$ , $\left[\varphi\right]_{w}\geq0$, were
used, this states an upper bound in the value of the critical $z$
exponent, namely $z=2$.

Hereafter, we particularize the condition (\ref{eq:II.3.1}) to the
case $z=2$ so that this becomes $N_{\partial_{i}}+\mathcal{N}_{\psi}+2\mathcal{N}_{F}\leq4$.
Since $\mathcal{N}_{\varphi}$ does not appear explicitly in this
inequality, any number of scalar $\varphi$ lines is allowed in a
given vertex $\mathcal{V}$ from the weighted renormalization viewpoint.
It should be noted also that the maximal number of spatial derivatives,
$N_{\partial_{i}}=4$, can only occur in pure scalar $\varphi$ vertices:
$\text{\ensuremath{\mathcal{V}}}\sim\partial^{i}\partial^{j}\bar{\varphi}\partial_{i}\partial_{j}\varphi,\,\partial^{i}\partial^{j}\bar{\varphi}\partial_{i}\partial_{j}\varphi\bar{\varphi}\varphi,\,\ldots$

The presence of spatial derivatives in the interaction vertices, in
particular, in genuine ones (i.e. vertices with more than two lines)
made the problem of finding the most general interaction Lagrangian
$\mathcal{L}_{int}$ extremely intricate in the component formalism.
Hence, for simplicity, we will seek $\mathcal{L}_{int}$ with the
following structure
\begin{equation}
\mathcal{L}_{int}=\bar{\psi}W_{1}\psi+\left[\frac{1}{2}\bar{\psi}W_{2}\bar{\psi}+W_{3}\bar{F}+h.c.\right]+V\bar{\psi}^{2}\psi^{2}+U,\label{eq:II.4}
\end{equation}
where $W_{i}$, $V$ and $U$ are functions of the scalar fields $\varphi$,
$\bar{\varphi}$. Note that due to the Hermiticity property of the
action (\ref{eq:II.1}), $W_{1}$, $V$ and $U$ have to be real operators,
whereas $W_{2}$ and $W_{3}$ complex ones. Furthermore, it is worthwhile
to note that in conformity with the $z=2$ wrc $\left(N_{\partial_{i}}+\mathcal{N}_{\psi}+2\mathcal{N}_{F}\leq4\right)$,
$W_{i}$ can contain at most two spatial derivatives $\left(N_{\partial_{i}}=2\right)$,
$U$ four spatial derivatives $\left(N_{\partial_{i}}=4\right)$,
and $V$ none $\left(N_{\partial_{i}}=0\right)$. Since we are interested
in adding higher spatial derivatives in the kinetic part of (\ref{eq:II.1}),
we shall look for expressions of the form $W_{1,2}\sim\Delta+\tilde{W}_{1,2}\left(\varphi,\bar{\varphi}\right)$,
$W_{3}\sim\Delta\left(\varphi\,\text{or}\,\bar{\varphi}\right)+\tilde{W}_{3}\left(\varphi,\bar{\varphi}\right)$,
and $U\sim\bar{\varphi}\Delta^{2}\varphi+\tilde{U}\left(\varphi,\bar{\varphi}\right)$,
where $\Delta=\partial^{i}\partial_{i}$ is the two-dimensional spatial
Laplace operator and where $\tilde{W}_{i}$, $\tilde{U}$ are functions
only of the scalar fields $\varphi,\bar{\varphi}$. With these considerations
in mind, one can easily compute the variation of $\mathcal{L}_{int}$
under the susy transformations (\ref{eq:II.2}). This can be cast
in the form
\begin{eqnarray}
\delta\mathcal{L}_{int} & = & \epsilon\psi\Sigma\bar{F}+\epsilon\bar{\psi}\left(\Pi+V\psi^{2}\right)\bar{F}-i\epsilon\gamma^{0}\bar{\psi}\left(\Sigma\partial_{0}\varphi+\Pi\partial_{0}\bar{\varphi}\right)-ia\epsilon\gamma^{i}\bar{\psi}\left(\Sigma\partial_{i}\varphi+\Pi\partial_{i}\bar{\varphi}\right)\nonumber \\
 &  & +\epsilon\bar{\psi}\psi^{2}\left(\partial_{\varphi}W_{1}-\partial_{\bar{\varphi}}\bar{W}_{2}\right)-iV\psi^{2}\left(\epsilon\gamma^{0}\bar{\psi}\partial_{0}\bar{\varphi}+a\epsilon\gamma^{i}\bar{\psi}\partial_{i}\bar{\varphi}\right)-\epsilon\psi\partial_{\varphi}U+h.c.,
\end{eqnarray}
where we have introduced the notation $\Sigma=W_{1}-\partial_{\varphi}W_{3}$
and $\Pi=W_{2}-\partial_{\bar{\varphi}}W_{3}$, and got rid of all
surface terms. From this result, it is clear that to respect susy
one must demand that $\Sigma=\Pi=V=U=0$. This in turn implies that
\begin{equation}
W_{1}=\frac{\partial W_{3}}{\partial\varphi}\qquad\qquad\qquad W_{2}=\frac{\partial W_{3}}{\partial\bar{\varphi}}.\label{eq:II.5}
\end{equation}
The operator $W_{3}$, on the other hand, is fully determined by imposing
the reality condition of $W_{1}$, $\partial W_{3}/\partial\varphi=\partial\overline{W_{3}}/\partial\bar{\varphi}$,
and the fact that its weight according to the $z=2$ wrc is $\left[W_{3}\right]_{w}=2$.
This means, as stated previously, that $W_{3}$ cannot contain more
than two spatial derivatives. After doing this, one gets
\begin{equation}
W_{3}=m\varphi+b\Delta\varphi+\sum_{p=1}^{\infty}g_{p}\varphi^{p+1}\bar{\varphi}^{p}.\label{eq:II.6}
\end{equation}
Notice, moreover, that susy excludes any possibility of introducing
explicitly four spatial derivatives $\left(N_{\partial_{i}}=4\right)$
in the off-shell formulation of the theory. Indeed, the desired Lifshitz
scalar operator $\bar{\varphi}\Delta^{2}\varphi$ appears only in
the bosonic sector of the theory after eliminating the auxiliary $F$
field (see Eq. (\ref{eq:II.9})). In the concluding part of this section,
we will confirm these results and generalize them in the Lifshitz
superfield formulation.

It is extremely important to expose the consistency of our model by
setting explicitly up its superalgebra at classical level. Hence,
in the balance of this section, we construct the Noether currents
(and their respective charges) associated with each symmetry of the
model under consideration and then we set up its superalgebra by using
the canonical (anti-)commutation relations.

For simplicity and without loss of generality, we truncate the series
in (\ref{eq:II.6}) at $p=1$. So, the interaction Lagrangian $\mathcal{L}_{int}$
of our Lifshitz-Wess-Zumino (L-WZ) model, as we shall call it, reads
\begin{align}
\mathcal{L}_{int} & =\bar{\psi}\left(m+b\Delta+2g\bar{\varphi}\varphi\right)\psi+\left[g\varphi^{2}\bar{\psi}^{2}+\left(m\varphi+b\Delta\varphi+g\varphi^{2}\bar{\varphi}\right)\bar{F}+h.c.\right].\label{eq:II.7}
\end{align}
Notice that setting $a\rightarrow1$ and $b\rightarrow0$ this theory
reduces to the usual relativistic Wess-Zumino model \cite{Gates et al (1983),Maluf=000026Silva (2013)}.

The L-WZ field equations which result from the principle of least
action, $\delta S=0$, are given by
\[
-\ddot{\varphi}+a^{2}\Delta\varphi+mF+b\Delta F+g\left(\varphi^{2}\bar{F}+2\bar{\varphi}\varphi F+2\bar{\varphi}\psi^{2}+2\varphi\bar{\psi}\psi\right)=0
\]
\begin{equation}
\left(i\cancel{\partial}_{0}+ia\cancel{\partial}_{i}+m+b\Delta\right)\psi+g\varphi\left(\varphi\bar{\psi}+2\bar{\varphi}\psi\right)=0\label{eq:II.8}
\end{equation}
\[
F+m\varphi+b\Delta\varphi+g\bar{\varphi}\varphi^{2}=0.
\]
In contrast with the relativistic Wess-Zumino model, it should be
noted that the auxiliary field equation contains the extra space differential
term $b\Delta\varphi$. The auxiliary field $F$ and its complex conjugate
$\bar{F}$, as mentioned earlier, play a leading role in the construction
of our Lifshitz susy field theory, since they introduce naturally
the right Lifshitz-operator, $\bar{\varphi}\triangle^{2}\varphi$,
in the bosonic sector of the theory. Indeed, after removing $F$ and
$\bar{F}$ from (\ref{eq:II.1}-\ref{eq:II.7}) by means of their
field equations, the bosonic part of the L-WZ Lagrangian reads
\begin{eqnarray}
-\mathcal{L}_{bos} & = & -\dot{\bar{\varphi}}\dot{\varphi}+\left(2mb-a^{2}\right)\bar{\varphi}\triangle\varphi+b^{2}\bar{\varphi}\triangle^{2}\varphi+m^{2}\bar{\varphi}\varphi+mg\left(\bar{\varphi}\varphi\right)^{2}\nonumber \\
 &  & +bg\bar{\varphi}\varphi\left(\varphi\triangle\bar{\varphi}+\bar{\varphi}\triangle\varphi\right)+g^{2}\left(\bar{\varphi}\varphi\right)^{3}.\label{eq:II.9}
\end{eqnarray}
It is important to point out that, as far as we know, this is the
first time this Lifshitz construction procedure in susy theories has
been proposed. This method, in particular, avoids the glaring inconsistencies
concerning the ill-definedness of the critical exponent $z$ as well
as the unnatural time-space mixing derivatives observed in \cite{Gomes et al (2014)}. 

Coming back to the problem of building up the superalgebra, we first
claim that it closes off-shell. Indeed, it is not hard to check that
the commutator of two susy transformations yields once again a symmetry
transformation, i.e. a linear asymmetric combination of time and space
transformations,
\begin{equation}
\left[\delta_{\epsilon_{2}},\delta_{\epsilon_{1}}\right]X=2i\left(\epsilon_{2}\gamma^{0}\epsilon_{1}\right)\partial_{0}X+2ai\left(\epsilon_{2}\gamma^{i}\epsilon_{1}\right)\partial_{i}X,\label{eq:II.10}
\end{equation}
where $X$ stands for the fields $\varphi$, $\psi_{\alpha}$ and
$F$. To prove (\ref{eq:II.10}) one has to make use of the Fierz
identity: $\chi_{\alpha}\left(\xi\eta\right)=-\xi_{\alpha}\left(\chi\eta\right)-\left(\xi\chi\right)\eta_{\alpha}$.
We should emphasize that this result and the others that we present
in the rest of this section, in particular (\ref{eq:II.18}), are
not only valid for the free theory, but also for the interaction one,
i.e. including the interaction Lagrangian (\ref{eq:II.7}). In other
words, we show explicitly that the implementation of the Lifshitz-like
operators $b\bar{\psi}\Delta\psi$ and $b^{2}\bar{\varphi}\Delta^{2}\varphi$,
this last by means of the auxiliary field $F$, in the fermionic and
bosonic sectors, respectively, does not spoil the susy algebra of
conventional (i.e. with $z=1$) Lorentz-violating supersymmetric theories
\cite{Berger=000026Kostelecky (2002)} at classical level.

According to the Noether theorem, it is not hard to show that the
components of the supercurrent of the L-WZ model associated with the
susy transformations (\ref{eq:II.2}) are given by
\begin{equation}
-J_{\alpha}^{0}=\bar{\psi}_{\alpha}\dot{\varphi}+\psi_{\alpha}\dot{\bar{\varphi}}+a\left(\psi\gamma^{0}\cancel{\partial}_{i}\right)_{\alpha}\bar{\varphi}+a\left(\bar{\psi}\gamma^{0}\cancel{\partial}_{i}\right)_{\alpha}\varphi+i\bar{F}\left(\gamma^{0}\psi\right)_{\alpha}+iF\left(\gamma^{0}\bar{\psi}\right)_{\alpha}\label{eq:II.11a}
\end{equation}
and 
\begin{equation}
-J_{\alpha}^{i}=\left(\psi\mathcal{S}^{i}\cancel{\mathcal{D}}\right)_{\alpha}\bar{\varphi}+\left(\bar{\psi}\mathcal{S}^{i}\cancel{\mathcal{D}}\right)_{\alpha}\varphi+i\bar{F}\left(\mathcal{S}^{i}\psi\right)_{\alpha}+iF\left(\mathcal{S}^{i}\bar{\psi}\right)_{\alpha},\label{eq:II.11b}
\end{equation}
where $\cancel{\mathcal{D}}_{\alpha\beta}=\left(\gamma^{0}\right)_{\alpha\beta}\partial_{0}+a\left(\gamma^{i}\right)_{\alpha\beta}\partial_{i}$
and $\mathcal{S}_{\alpha\beta}^{i}=a\left(\gamma^{i}\right)_{\alpha\beta}-ibC_{\alpha\beta}\overleftrightarrow{\partial}^{i}$.
Using the field equations (\ref{eq:II.8}), one can verify the conservation
of the supercurrent $J_{\alpha}^{\mu}$, namely $\partial_{\mu}J_{\alpha}^{\mu}=0$.
As in conventional supersymmetric theories, it follows that the conserved
supercharge $Q_{\alpha}=\int d^{2}\mathbf{x}J_{\alpha}^{0}$ generates
the susy transformations (\ref{eq:II.2}). That is,
\begin{equation}
\left[\epsilon^{\alpha}Q_{\alpha},\,X\right]=-i\delta X.\label{eq:II.12}
\end{equation}
This relation may be checked out by using the classical field equations
(\ref{eq:II.8}) and the canonical (anti-)commutators for the fields,
\begin{equation}
\left[\varphi\left(\mathbf{x}\right),\dot{\bar{\varphi}}\left(\mathbf{y}\right)\right]=i\delta_{\mathbf{x},\mathbf{y}},\quad\left[\bar{\varphi}\left(\mathbf{x}\right),\dot{\varphi}\left(\mathbf{y}\right)\right]=i\delta_{\mathbf{x},\mathbf{y}},\quad\left[\psi_{\alpha}\left(\mathbf{x}\right),\,\bar{\psi}_{\beta}\left(\mathbf{y}\right)\right]=\left(\gamma^{0}\right)_{\alpha\beta}\delta_{\mathbf{x},\mathbf{y}},\label{eq:II.13}
\end{equation}
where we have omitted for simplicity the equal time variable $t$
in the argument of the fields and used $\delta_{\mathbf{x},\mathbf{y}}$
to represent the two-dimensional Dirac delta function $\delta^{2}\left(\mathbf{x}-\mathbf{y}\right)$. 

In order to assemble the susy algebra we follow the same technique
as in usual susy theories \cite{Terning (2009)} . Namely, we employ
the Jacobi identity,
\begin{equation}
\left[\left[A,B\right],C\right]=\left[A,\left[B,C\right]\right]-\left[B,\left[A,C\right]\right],\label{eq:II.14}
\end{equation}
with $A=\epsilon_{2}Q$, $B=\epsilon_{1}Q$ and $C=X$. After simplifying
this with the help of the formulas (\ref{eq:II.10}) and (\ref{eq:II.12}),
one gets
\begin{equation}
\left[\left[\epsilon_{2}Q,\epsilon_{1}Q\right],\,X\right]=-2i\left(\epsilon_{2}\gamma^{0}\epsilon_{1}\right)\partial_{0}X-2ai\left(\epsilon_{2}\gamma^{i}\epsilon_{1}\right)\partial_{i}X.\label{eq:II.15}
\end{equation}
At this point it is important to recognize that the L-WZ action (\ref{eq:II.1})
is invariant under other symmetries. Invariance under time translations,
$\delta_{\tau}X=\tau\partial_{0}X$, gives rise to a conserved current
$\tilde{J}^{\mu}$ whose components are
\begin{equation}
\tilde{J}^{0}=\dot{\bar{\varphi}}\dot{\varphi}+a^{2}\partial^{i}\bar{\varphi}\partial_{i}\varphi-ia\bar{\psi}\cancel{\partial}_{i}\psi-m\bar{\psi}\psi-b\bar{\psi}\Delta\psi+\bar{F}F-g\left(\varphi^{2}\bar{\psi}^{2}+\bar{\varphi}^{2}\psi^{2}+2\varphi\bar{\varphi}\bar{\psi}\psi\right)\label{eq:II.16a}
\end{equation}
and
\begin{equation}
\tilde{J}^{i}=-a^{2}\left(\dot{\bar{\varphi}}\partial^{i}\varphi+\dot{\varphi}\partial^{i}\bar{\varphi}\right)+ia\bar{\psi}\gamma^{i}\dot{\psi}+b\left(\bar{F}\overleftrightarrow{\partial}^{i}\dot{\varphi}+F\overleftrightarrow{\partial}^{i}\dot{\bar{\varphi}}+\bar{\psi}\overleftrightarrow{\partial}^{i}\dot{\psi}\right).\label{eq:II.16b}
\end{equation}
Of course, in this case, the conserved charge $H=\int d^{2}\mathbf{x}\tilde{J}^{0}$
turns out to be the Hamiltonian of the Lifshitz-Wess-Zumino model.
By using the canonical relations (\ref{eq:II.13}), one may show that
$H$ is in effect the generator of the time translations: $\left[\tau H,\,X\right]=-i\delta_{\tau}X$. 

Invariance under spatial translations, $\delta_{s}X=s^{k}\partial_{k}X$,
gives rise to two conserved currents $T_{\,\,k}^{\mu}$ (note that
in planar physics there are only two spatial directions) whose components
are
\begin{equation}
T_{\,\,k}^{0}=\partial_{k}\bar{\varphi}\dot{\varphi}+\dot{\bar{\varphi}}\partial_{k}\varphi+i\bar{\psi}\gamma^{0}\partial_{k}\psi\label{eq:II.17a}
\end{equation}
and 
\begin{eqnarray}
T_{\,\,k}^{i} & = & -a^{2}\left(\partial_{k}\bar{\varphi}\partial^{i}\varphi+\partial^{i}\bar{\varphi}\partial_{k}\varphi\right)+ia\bar{\psi}\gamma^{i}\partial_{k}\psi-b\left(\partial_{k}\varphi\partial^{i}\bar{F}+\partial^{i}\varphi\partial_{k}\bar{F}\right.\nonumber \\
 &  & \left.+\partial_{k}\bar{\varphi}\partial^{i}F+\partial^{i}\bar{\varphi}\partial_{k}F+\partial_{k}\bar{\psi}\partial^{i}\psi+\partial^{i}\bar{\psi}\partial_{k}\psi\right)-\delta_{k}^{i}\mathcal{L}.\label{eq:17b}
\end{eqnarray}
The two conserved charges $P_{k}=\int d^{2}\mathbf{x}T_{\,\,k}^{0}$
define the momentum vector of the system. Once again it is easy to
see that $P_{k}$ are the generators of space translations: $\left[s^{k}P_{k},\,X\right]=-i\delta_{s}X$.

With these results at hand, we eliminate the time and space derivatives
from (\ref{eq:II.15}) in terms of the $H$ and $P_{k}$ commutators.
In doing this, we finally obtain for the susy algebra
\begin{equation}
\left\{ Q_{\alpha},\,Q_{\beta}\right\} =2\left(\gamma^{0}\right)_{\alpha\beta}H+2a\left(\gamma^{i}\right)_{\alpha\beta}P_{i}.\label{eq:II.18}
\end{equation}
This anticommutator of two supercharges is the three-dimensional version
of that found in \cite{Berger=000026Kostelecky (2002)} regarding
the violation of the Lorentz symmetry in the conventional four-dimensional
Wess-Zumino model. Note however that in the present investigation
we are considering Lorentz-violation operators with higher space derivatives,
i.e. Lifshitz-like operators, a marked difference with respect to
\cite{Berger=000026Kostelecky (2002)}.

For completeness, we compute the Noether current $J_{R}^{\mu}$ associated
to the invariance of the L-WZ action (\ref{eq:II.1}) under the rotation
group $SO\left(2\right)$. The rotational transformations that leave
the action invariant are given by
\begin{equation}
\delta_{\theta}\varphi=-i\theta\hat{L}\varphi\qquad\delta_{\theta}F=-i\theta\hat{L}F\qquad\delta_{\theta}\psi=-i\theta\hat{L}\psi-i\theta\Sigma\psi,\label{eq:II.19}
\end{equation}
where $\hat{L}=i\left(x^{2}\partial_{1}-x^{1}\partial_{2}\right)$
denotes the angular momentum generator and $\Sigma=-i\left[\gamma^{1},\,\gamma^{2}\right]/4$
the spinor generator. One can show by using the Noether's method that
the components of the conserved current $J_{R}^{\mu}$ are
\begin{equation}
J_{R}^{0}=x^{2}T_{\,\,1}^{0}-x^{1}T_{\,\,2}^{0}+\bar{\psi}\gamma^{0}\Sigma\psi\label{eq:II.20a}
\end{equation}
and
\begin{equation}
J_{R}^{i}=x^{2}T_{\,\,1}^{i}-x^{1}T_{\,\,2}^{i}+a\bar{\psi}\gamma^{i}\Sigma\psi-ib\bar{\psi}\Sigma\partial^{i}\psi+ib\partial^{i}\bar{\psi}\Sigma\psi.\label{eq:II.20b}
\end{equation}
The Noether charge $\mathcal{J}=\int d^{2}\mathbf{x}J_{R}^{0}$ corresponds
to the angular momentum of the system and its conservation is guaranteed
by $\partial_{\mu}J_{R}^{\mu}=0$. Note that $\mathcal{J}$ is, as
should be, the generator of the rotational transformations (\ref{eq:II.19}),
i.e., $\left[\theta\mathcal{J},\,X\right]=-i\delta_{\theta}X$.

We close this section by rephrasing our model in what we call the
Lifshitz superspace, a natural modification of the conventional one
to treat susy theories with Lifshitz-like operators. For this purpose,
as in the usual case, we first compact the fields $\varphi$, $\psi_{\alpha}$
and $F$ into a scalar superfield $\Phi$,
\begin{equation}
\Phi\left(x^{0},\,x^{i},\,\theta\right)=\varphi+\theta^{\alpha}\psi_{\alpha}-\theta^{2}F.\label{eq:II.21}
\end{equation}
Next, since the time and space coordinates are weighted differently
in Lifshitz field theories, we split the usual susy covariant derivative
$D_{\alpha}=\partial_{\alpha}+i\left(\gamma^{\mu}\right)_{\alpha\beta}\theta^{\beta}\partial_{\mu}$
into a time 'covariant' derivative $D_{t\alpha}$ and a space 'covariant'
derivative $D_{s\alpha}$:
\begin{equation}
2D_{t\alpha}=\partial_{\alpha}+2i\left(\gamma^{0}\right)_{\alpha\beta}\theta^{\beta}\partial_{0}\qquad\qquad2D_{s\alpha}=\partial_{\alpha}+2ia\left(\gamma^{i}\right)_{\alpha\beta}\theta^{\beta}\partial_{i},\label{eq:II.22a}
\end{equation}
where the parameter $a$ is essential for counterbalancing the weight
of the spatial $\partial_{i}$ derivative compared with the weight
of the time $\partial_{0}$ derivative. This coincides with the one
introduced in (\ref{eq:II.1}) and weighs $\left[a\right]_{w}=z-1$=1,
while $\left[\theta_{\alpha}\right]_{w}=-z/2=-1$. Within this Lifshitz
superspace formulation, the term 'covariant' must be taken with great
care, for these susy derivatives are covariant regarding the time
and space supercharges $Q_{t\alpha}$ and $Q_{s\alpha}$ defined in
Appendix \ref{sec:Appendix.A}, but not with respect to the net supercharge
$\mathcal{Q}_{\alpha}=Q_{t\alpha}+Q_{s\alpha}$ that realizes the
susy algebra (\ref{eq:II.18}). The net covariant derivative $\mathcal{D}_{\alpha}$
which anticommutes with $\mathcal{Q}_{\alpha}$ (see Appendix \ref{sec:Appendix.A}
for details) is given by
\begin{equation}
\mathcal{D}_{\alpha}=D_{t\alpha}+D_{s\alpha}=\partial_{\alpha}+i\left(\gamma^{0}\right)_{\alpha\beta}\theta^{\beta}\partial_{0}+ia\left(\gamma^{i}\right)_{\alpha\beta}\theta^{\beta}\partial_{i}.\label{eq:II.22b}
\end{equation}
Note that this weighted covariant derivative $\mathcal{D}_{\alpha}$
becomes the usual one taking $a\rightarrow1$.

In terms of these superobjects, the superfield counterpart of the
L-WZ action can be cast in the form
\begin{equation}
S=\int d^{5}\tilde{z}\left\{ -\frac{1}{2}\mathcal{D}^{\alpha}\bar{\Phi}\mathcal{D}_{\alpha}\Phi+\frac{4b}{a^{2}}D_{s}^{2}\bar{\Phi}D_{s}^{2}\Phi+m\bar{\Phi}\Phi+\frac{g}{2}\left(\bar{\Phi}\Phi\right)^{2}\right\} ,\label{eq:II.23}
\end{equation}
where $d^{5}\tilde{z}=dtd^{2}xd^{2}\theta$ is the superspace measure.
By carrying out explicitly the Grassmann integration or by doing this
with the help of the projection techniques described in Appendix \ref{sec:Appendix.A},
it is straightforward to show that this superaction reduces to the
component one (\ref{eq:II.1}) with $\mathcal{L}_{int}$ given by
(\ref{eq:II.7}). 

It is interesting to see that the higher space derivative term (apparently
not covariant in the entire Lifshitz superspace by the presence of
the space derivative $D_{s}$) is indeed covariant up to surface terms.
To see this more closely, we integrate it by parts
\begin{eqnarray}
\int dtd^{2}x\left.4D_{s}^{2}\left[\left(\frac{4b}{a^{2}}\right)D_{s}^{2}\bar{\Phi}D_{s}^{2}\Phi\right]\right| & = & \int dtd^{2}x\left(\frac{8b}{a^{2}}\right)\left.D_{s}^{2}\left[D_{s}^{\alpha}\left(-\bar{\Phi}\overleftrightarrow{D}_{s\alpha}D_{s}^{2}\Phi\right)+2\bar{\Phi}\left(D_{s}^{2}\right)^{2}\Phi\right]\right|\nonumber \\
 & = & \int dtd^{2}x\left.4D_{s}^{2}\left[b\bar{\Phi}\Delta\Phi\right]\right|=\int dtd^{2}xd^{2}\theta\left(b\bar{\Phi}\Delta\Phi\right),\label{eq:II.24}
\end{eqnarray}
where in the last line we have ignored the surface space term $D_{s}^{2}D_{s}^{\alpha}\left(\cdots\right)\sim\partial_{i}\left(\cdots\right)$
and used the identity $\left(D_{s}^{2}\right)^{2}=a^{2}\Delta/4$.
Since $\partial_{i}$ is a covariant derivative in the entire Lifshitz
superspace, i. e., $\left[\partial_{i},\mathcal{Q}_{\alpha}\right]=0$,
the term $\bar{\Phi}\Delta\Phi$ is manifest covariant.

We now reproduce and generalize our $z=2$ component results in the
Lifshitz superspace formulation. Let $\mathscr{V}_{s}$ be a supervertex
of the form $\mathscr{V}_{s}=g\,\partial_{i}^{N_{\partial_{i}}}\left(\bar{\Phi}\Phi\right)^{\mathcal{N}_{\Phi}/2}$,
where $g$ represents a coupling constant and $\mathcal{N}_{\Phi}$
the total number of scalar $\Phi$ superfields. This vertex is supersymmetric
by construction, for it involves just susy covariant objects. Note
also that the spatial rotational $SO\left(2\right)$ symmetry restricts
$N_{\partial_{i}}$ to zero or even values $\left(N_{\partial_{i}}=0,\,2,\,4,\ldots\right)$
with all the $\partial_{i}$ completely contracted, while the rigid
phase $U\left(1\right)$ symmetry, $\Phi\rightarrow e^{i\alpha}\Phi$,
restricts $\mathcal{N}_{\Phi}$ to only even values $\left(\mathcal{N}_{\Phi}=2,\,4,\ldots\right)$.
The wrc applied to this kind of vertex states that $\mathscr{V}_{s}$
will be renormalizable by weighted power counting iff the supercondition
\begin{equation}
N_{\partial_{i}}+\frac{\left(2-z\right)}{2}\mathcal{N}_{\Phi}\leq2\label{eq:II.25}
\end{equation}
were satisfied. This outcome is derived from the weighted analysis
of $\mathscr{V}_{s}$ and $\Phi$, and from demanding $\left[g\right]_{w}\geq0$.
The weights of $\mathscr{V}_{s}$ and $\Phi$ are found as follows.
As the supermeasure $d^{5}\tilde{z}$ in (\ref{eq:II.23}) weighs
$\left[d^{5}\tilde{z}\right]_{w}=-2$, for whatever value of $z$,
it follows that the weight of $\mathscr{V}_{s}$, as a part of the
superlagrangian $\mathscr{L}_{s}$, must be $\left[\mathscr{V}_{s}\right]_{w}=\left[\mathscr{L}_{s}\right]_{w}=2$.
On the other hand, the weight of the scalar $\Phi$ superfield turns
out to be $\left[\Phi\right]_{w}=\left[\varphi\right]_{w}=\left(2-z\right)/2$,
as seen directly from (\ref{eq:II.21}) or indireclty from the kinetic
part of (\ref{eq:II.23}). 

The weighted renormalization supercondition (\ref{eq:II.25}) reproduces
the well-known results for $z=1$ and our component results for $z=2$.
Indeed, as is well known, the conventional power counting renormalization
in the three dimensional superfield formulation of the Wess-Zumino
model allows vertices with at most four scalar superfields $(\mathcal{N}_{\Phi}\leq4)$
and without any explicit spatial derivative $\left(N_{\partial_{i}}=0\right)$.
This result follows at once from (\ref{eq:II.25}) taking $z=1$,
as expected. In the $z=2$ case, one can see that this condition becomes
simply $N_{\partial_{i}}\leq2$, confirming so our previous component
results, namely a vertex can contain at most two spatial derivatives
and any number of scalar $\Phi$ lines. It is not difficult to prove
that the component interaction Lagrangian $\mathcal{L}_{int}$ constructed
in (\ref{eq:II.4}), with $W_{i}$ given in (\ref{eq:II.5}) and (\ref{eq:II.6})
and with $V=U=0$, corresponds to the superlagrangian $\mathscr{L}_{s,int}$
given by
\begin{equation}
\mathscr{L}_{s,int}=m\bar{\Phi}\Phi+b\bar{\Phi}\Delta\Phi+\sum_{p=1}^{\infty}\frac{g_{p}}{p+1}\left(\bar{\Phi}\Phi\right)^{p+1}.\label{eq:II.26}
\end{equation}
Clearly, the superfield $z=2$ analysis carried out here permits to
generalize (\ref{eq:II.26}) by adding to it spatial derivative genuine
vertices, i.e vertices of the type $\mathscr{V}_{s}\sim\bar{\Phi}\Phi\partial^{i}\bar{\Phi}\partial_{i}\Phi,\ldots,\left(\partial_{i}\right)^{2}\left(\bar{\Phi}\Phi\right)^{\mathcal{N}_{\Phi}/2}$,
with $\mathcal{N}_{\Phi}>2$. 

We conclude from our research that the wrc is mandatory in the construction
of well-defined susy Lifshitz field theories. In Appendix \ref{sec:Appendix.B},
we will illustrate how this Lifshitz superfield method works by implementing
Lifshitz-like operators in the four dimensional Wess-Zumino model.

\section{The effective potential to one-loop order\label{sec:Sec3}}

This section is devoted to investigate the vacuum quantum effects
of the Lifshitz operators implemented in the conventional Wess-Zumino
model. For this purpose, we shall compute the one-loop effective potential
of the L-WZ model (\ref{eq:II.1}) with $\mathcal{L}_{int}$ defined
in (\ref{eq:II.7}). Moreover, we shall take advantage of this calculation
to examine the improvement of the ultraviolet behavior in this kind
of theory. As is well known, up to a spacetime volume $v_{3}=\int d^{3}x$,
the zero-loop potential $V_{0}$ is the negative of the classical
action evaluated at the position-independent fields $\varphi\left(x\right)=\varphi_{0}=\left(\sigma_{1}+i\pi_{1}\right)/\sqrt{2}$
, $F(x)=f_{0}=\left(\sigma_{2}+i\pi_{2}\right)/\sqrt{2}$ and $\psi_{\alpha}\left(x\right)=0$.
In doing this, one gets
\begin{eqnarray}
V_{0} & = & -\frac{1}{2}\left[\sigma_{2}^{2}+\pi_{2}^{2}+2m\left(\sigma_{1}\sigma_{2}+\pi_{1}\pi_{2}\right)+g\left(\sigma_{1}^{2}+\pi_{1}^{2}\right)\left(\sigma_{1}\sigma_{2}+\pi_{1}\pi_{2}\right)\right]\nonumber \\
 & = & \frac{m^{2}}{2}\left(\sigma_{1}^{2}+\pi_{1}^{2}\right)+\frac{mg}{2}\left(\sigma_{1}^{2}+\pi_{1}^{2}\right)^{2}+\frac{g^{2}}{8}\left(\sigma_{1}^{2}+\pi_{1}^{2}\right)^{3},\label{eq:III.1}
\end{eqnarray}
where in the last equality we have eliminated the real auxiliary fields
$\sigma_{2}$ and $\pi_{2}$ by means of their field equations $\partial V_{0}/\partial\sigma_{2}=0$
and $\partial V_{0}/\partial\pi_{2}=0$. It is not hard to see that
at classical level the theory exhibits two phases in regard to the
spontaneous breaking of the global phase $U(1)$ symmetry group: $\sigma_{i}'+i\pi_{i}'=\exp\left(i\alpha\right)\left(\sigma_{i}+i\pi_{i}\right)$.
These two phases are dictated by the sign of the order parameter $\xi=m/g$.
In fact, by analyzing the minima of the classical potential (\ref{eq:III.1}),
we conclude that whether $\xi\geq0$, the $U(1)$ symmetry is preserved,
since the vacuum state is unique and it corresponds to the trivial
one $\sigma_{1}=0=\pi_{1}$. On the other hand, if $\xi<0$, the $U(1)$
symmetry is spontaneously broken, for in addition to the trivial vacuum
state, there is a manifold of non-trivial vacuum states defined by
$\sigma_{1}^{2}+\pi_{1}^{2}=2\left|\xi\right|$. Note that supersymmetry
in both phases remains intact due to the vanishing of the vacuum energy.
In what follows, we confine our attention to the case $\xi>0$, considering
$m>0$ and $g>0$.

In order to compute the one-loop contribution for the effective potential
we employ the steepest-descent method \cite{Itzykson=000026Zuber (1980)}.
According to this method, the one-loop contribution becomes
\begin{equation}
V_{1}=-\frac{i}{2v_{3}}\ln\det\left(\mathcal{Q}^{2}-4\bar{\mathcal{R}}\mathcal{R}\right)+\frac{i}{2v_{3}}\ln\det\left(\mathcal{Q}_{\alpha\beta}^{2}-g^{2}\left|\varphi_{0}\right|^{4}C_{\alpha\beta}\right),\label{eq:III.2}
\end{equation}
where, defining the field-dependent mass $M=m+2g\left|\varphi_{0}\right|^{2}$,
\begin{equation}
\mathcal{Q}=\partial_{0}^{2}+\left(a^{2}-2b\,M\right)\Delta-b^{2}\Delta^{2}-M^{2}+2g\left(\varphi_{0}\bar{f}_{0}+\bar{\varphi}_{0}f_{0}\right)-g^{2}\left|\varphi_{0}\right|^{4},\label{eq:III.3.1}
\end{equation}
\begin{equation}
\mathcal{R}=g\varphi_{0}f_{0}-g\,M\varphi_{0}^{2}-gb\varphi_{0}^{2}\Delta,\label{eq:III.3.2}
\end{equation}
\begin{equation}
\mathcal{Q}_{\alpha\beta}=i\left(\gamma^{0}\right)_{\alpha\beta}\partial_{0}+ia\left(\gamma^{i}\right)_{\alpha\beta}\partial_{i}+C_{\alpha\beta}\left(M+b\Delta\right).\label{eq:III.3.3}
\end{equation}
Using the $\zeta$-functional method \cite{Hawking (1977)} for solving
the functional determinants in (\ref{eq:III.2}), one gets{\small{}
\begin{eqnarray}
V_{1} & = & -\frac{i}{2}\int\frac{d^{3}k}{\left(2\pi\right)^{3}}\left\{ \ln\left[-k_{0}^{2}+\left(a^{2}-2b\mu_{1}\right)\mathbf{k}^{2}+b^{2}\mathbf{k}^{4}+\mu_{1}^{2}-g\sigma_{1}\sigma_{2}\right]+\ln\left[-k_{0}^{2}+\left(a^{2}-2b\mu_{2}\right)\mathbf{k}^{2}\right.\right.\nonumber \\
 &  & \left.\left.+b^{2}\mathbf{k}^{4}+\mu_{2}^{2}-3g\sigma_{1}\sigma_{2}\right]\right\} +\frac{i}{2}\int\frac{d^{3}k}{\left(2\pi\right)^{3}}\left\{ \ln\left[-k_{0}^{2}+\left(a^{2}-2b\mu_{1}\right)\mathbf{k}^{2}+b^{2}\mathbf{k}^{4}+\mu_{1}^{2}\right]\right.\nonumber \\
 &  & \left.+\ln\left[-k_{0}^{2}+\left(a^{2}-2b\mu_{2}\right)\mathbf{k}^{2}+b^{2}\mathbf{k}^{4}+\mu_{2}^{2}\right]\right\} ,\label{eq:III.4}
\end{eqnarray}
}where $\mu_{i}$ are the field-dependent masses $\mu_{i}=m+\left(2i-1\right)g\sigma_{1}^{2}/2$,
with $i=1,\,2$. With respect to this result, some comments are pertinent.
First, for simplicity and without loss of generality, we have set
$\pi_{i}=0$ in (\ref{eq:III.2}). This is always possible owing to
the global $U(1)$ symmetry of the effective potential. Second, in
the analysis of the cancellation of infinities that will be carried
out below, it is important to keep in mind that the first integral
becomes from the bosonic determinant, whereas the second one (that
with positive sign) from the fermionic one.

We compute the integrals in (\ref{eq:III.4}) by using the formula{\small{}
\begin{eqnarray}
\int\frac{d^{3}k}{\left(2\pi\right)^{3}}\ln\left[-k_{0}^{2}+x^{2}\mathbf{k}^{2}+y^{2}\mathbf{k}^{4}+z^{2}\right] & = & \frac{i}{32\pi y^{3}}\left[2y^{2}z^{2}-2x^{2}yz+\left(x^{4}-4y^{2}z^{2}\right)\ln\left(x^{2}+2yz\right)\right]\nonumber \\
 &  & +C_{1\Lambda}\left(y\right)z^{2}+C_{2\Lambda}\left(x,y\right),\label{eq:III.5a}
\end{eqnarray}
}where $C_{i}$ are infinite constants given by
\begin{equation}
C_{1}\left(x,y\right)=\frac{i}{2}\int\frac{d^{2}\mathbf{k}}{\left(2\pi\right)^{2}}\frac{1}{\sqrt{x^{2}\mathbf{k}^{2}+y^{2}\mathbf{k}^{4}}}+\frac{i}{8\pi y}\ln x^{2}\xrightarrow{\Lambda\rightarrow\infty}\frac{i}{4\pi y}\ln\left(2y\Lambda\right)\label{eq:III.5b}
\end{equation}
\begin{eqnarray}
C_{2}\left(x,y\right) & = & i\int\frac{d^{2}\mathbf{k}}{\left(2\pi\right)^{2}}\sqrt{x^{2}\mathbf{k}^{2}+y^{2}\mathbf{k}^{4}}-\frac{ix^{4}}{32\pi y^{3}}\ln x^{2}\nonumber \\
 & \xrightarrow{\Lambda\rightarrow\infty} & \frac{i}{64\pi y^{3}}\left[x^{4}\left(1-4\ln\left(2y\Lambda\right)\right)+8x^{2}y^{2}\Lambda^{2}+8y^{4}\Lambda^{4}\right]\label{eq:III.5c}
\end{eqnarray}
Here $\Lambda$ represents an UV cutoff in the two-dimensional momentum
space. Note also that the infinite constant $C_{1}$, for large $\Lambda$,
turns out to be independent of the $x-$parameter, yet $C_{1}$ is
a function of $x$ and $y$ for finite values of $\Lambda$. Adding
the result of (\ref{eq:III.4}) to (\ref{eq:III.1}), with $\pi_{i}=0$,
and then renormalizing it as described below, one gets the following
expression for the renormalized 1-loop effective potential
\begin{eqnarray}
\frac{V_{eff}}{m^{3}} & = & -\frac{1}{2}\left(\sigma_{2}^{2}+2\sigma_{1}\sigma_{2}+g\sigma_{1}^{3}\sigma_{2}\right)+\frac{1}{64\pi b^{3}}\Biggl\{2b\left(a^{2}-2b\mu_{1}\right)M_{1}+2b\left(a^{2}-2b\mu_{2}\right)M_{2}\nonumber \\
 &  & +a^{2}\left(a^{2}-4b\mu_{1}\right)\ln\left(1-\frac{2bM_{1}}{a^{2}}\right)+a^{2}\left(a^{2}-4b\mu_{2}\right)\ln\left(1-\frac{2bM_{2}}{a^{2}}\right)\nonumber \\
 &  & +4gb^{2}\sigma_{1}\sigma_{2}\left[\ln\left(1-\frac{2bM_{1}}{a^{2}}\right)+3\ln\left(1-\frac{2bM_{2}}{a^{2}}\right)+24\pi b\eta^{2}-2\right]\Biggr\},\label{eq:III.6a}
\end{eqnarray}
where $M_{i}=\mu_{i}-\sqrt{\mu_{i}^{2}-\left(2i-1\right)g\sigma_{1}\sigma_{2}}$
and $\eta$ is a renormalization point, defined by the equation
\begin{equation}
\left.\frac{1}{m^{3}}\frac{\partial^{2}V_{eff}}{\partial\sigma_{1}\partial\sigma_{2}}\right|_{\sigma_{1}=\eta,\,\sigma_{2}=0}=-1.\label{eq:III.6b}
\end{equation}
 On the right-hand side of (\ref{eq:III.6a}), we have made all quantities
dimensionless by rescaling these in terms of the mass $m$ parameter,
i. e., $\sigma_{1}\rightarrow m^{1/2}\sigma_{1}$, $\sigma_{2}\rightarrow m^{3/2}\sigma_{2}$,
$b\rightarrow m^{-1}b$. Notice in particular that $\mu_{i}$ in this
equation stands for $\mu_{i}=1+\left(2i-1\right)g\sigma_{1}^{2}/2.$ 

Before analyzing the minima of the effective potential, some remarks
are in order in connection with the renormalization procedure used
above. Note firstly that there was a complete cancellation of the
$C_{2}$ infinities between the bosonic and fermionic contributions.
This cancellation is essential for the UV improvement of the theory,
since the conventional (i.e. without Lifshitz operators) three-dimensional
Wess-Zumino model \cite{Burgess (1983),Gallegos et al (2013)} contains
only logarithmic and linear divergences at one-loop order and $C_{2}$
in (\ref{eq:III.5c}) contains field-dependent quadratic divergences.
The quartic divergences that might appear in the unrenormalized effective
potential $V_{eff}$ via (\ref{eq:III.5c}) if this cancellation fails
are of course matterless, for these are field-independent and might
eventually be absorbed by introducing a 'cosmological' constant in
the Lagrangian of the model. On the other hand, the cancellation of
the $C_{1}$ infinities (logarithmic divergences) was partial. This
residual susy divergence has been absorbed by adding a mass-type counterterm,
$A\sigma_{1}\sigma_{2}$, in the definition of the 1-loop effective
potential: $m^{-3}V_{eff}=V_{0}+V_{1}+A\sigma_{1}\sigma_{2}$. By
imposing the renormalization condition (\ref{eq:III.6b}), it is easy
to show that $A=\frac{g}{2\pi b}\left[3\pi b\eta^{2}+\ln\left(\frac{2b\Lambda}{a}\right)\right]$.

Let us now examine the stationary conditions of the renormalized 1-loop
effective potential (\ref{eq:III.6a}). Defining $\mathcal{F}_{ab}\left(x\right)=\ln\left(1-2bx/a^{2}\right)$,
they can be cast in the form
\begin{eqnarray}
\frac{1}{m^{3}}\frac{\partial V_{eff}}{\partial\sigma_{1}} & = & -\sigma_{2}-\frac{3}{2}g\left(\sigma_{1}^{2}-\eta^{2}\right)\sigma_{2}-\frac{g\sigma_{1}}{8\pi b}\left(M_{1}+3M_{2}\right)\nonumber \\
 &  & -\frac{ga^{2}}{16\pi b^{2}}\left(\sigma_{1}-\frac{b\sigma_{2}}{a^{2}}\right)\left[\mathcal{F}_{ab}\left(M_{1}\right)+3\mathcal{F}_{ab}\left(M_{2}\right)\right]=0\label{eq:III.7a}
\end{eqnarray}
and
\begin{equation}
\frac{1}{m^{3}}\frac{\partial V_{eff}}{\partial\sigma_{2}}=-\sigma_{2}-\sigma_{1}-\frac{g}{2}\sigma_{1}^{3}+\frac{g\sigma_{1}}{16\pi b}\left[\mathcal{F}_{ab}\left(M_{1}\right)+3\mathcal{F}_{ab}\left(M_{2}\right)+24\pi b\eta^{2}\right]=0.\label{eq:III.7b}
\end{equation}
This pair of coupled equations, in principle, can be solved by first
finding $\sigma_{2}=\sigma_{2}\left(\sigma_{1}\right)$ from (\ref{eq:III.7b})
and then plugging it back into (\ref{eq:III.7a}) to obtain the stationary
point: $\sigma_{1}=\tilde{\sigma}_{1}$ and $\tilde{\sigma}_{2}=\sigma_{2}\left(\tilde{\sigma}_{1}\right)$.
This procedure, however, is impracticable due to the intricate form
of the stationary equations and the field dependence of $M_{i}=M_{i}\left(\sigma_{1},\sigma_{2}\right)$.
Despite this fact, these equations provide relevant information for
the study of spontaneous (super)symmetry breaking \cite{Alvarez-Gaume (1982)}.
To see this, we must first observe that the condition
\begin{equation}
-\sigma_{1}-\frac{g}{2}\sigma_{1}^{3}+\frac{3}{2}g\eta^{2}\sigma_{1}=0,\label{eq:III.8}
\end{equation}
which becomes from (\ref{eq:III.7b}) taking $\sigma_{2}=0$, has
one real root at $\sigma_{1}=0$ for $\eta\leq\eta_{c}$ and three
real roots at $\sigma_{1}=0$ and $\sigma_{1}=\pm\sqrt{\left(3\eta^{2}g-2\right)/g}$
for $\eta>\eta_{c}$, where $\eta_{c}=\sqrt{2/3g}$ is a critical
value of $\eta$. Next let us denote any of these roots by $\tilde{\sigma}_{1}$
and note that $\sigma_{2}\left(\tilde{\sigma}_{1}\right)=0$. Since
(\ref{eq:III.7a}) is also satisfied at $\sigma_{2}=0$, we conclude
that the field configuration defined by $\sigma_{1}=\tilde{\sigma}_{1}$
and $\sigma_{2}=0$ is a stationary one. We notice, furthermore, that
the effective potential $V_{eff}$ vanishes at this stationary point,
i. e., $V_{eff}\left(\tilde{\sigma}_{1},\,\sigma_{2}\left(\tilde{\sigma}_{1}\right)=0\right)=0$.
It follows from this fact and the positivity condition of the effective
potential, $V_{eff}\left(\sigma_{1},\sigma_{2}\left(\sigma_{1}\right)\right)\geq0$,
that this stationary configuration is really an absolute minimum (with
zero energy) and so susy remains unbroken at one-loop order. It is
worthwhile to mention that the positivity condition of the energy
in conventional susy theories holds in this kind of theory and is
indeed assured by the susy algebra (\ref{eq:II.18}). One can state
explicitly this condition by rewritten the effective potential (\ref{eq:III.6a})
with the aid of (\ref{eq:III.7b}) in the form
\begin{eqnarray}
\frac{V_{eff}}{m^{3}} & = & \frac{\sigma_{2}^{2}}{2}+\frac{1}{64\pi b^{3}}\left[2b\left(a^{2}-2b\mu_{1}\right)M_{1}+2b\left(a^{2}-2b\mu_{2}\right)M_{2}\right.\nonumber \\
 &  & \left.+a^{2}\left(a^{2}-4b\mu_{1}\right)\mathcal{F}_{ab}\left(M_{1}\right)+a^{2}\left(a^{2}-4b\mu_{2}\right)\mathcal{F}_{ab}\left(M_{2}\right)-8gb^{2}\sigma_{1}\sigma_{2}\right]
\end{eqnarray}
and noting that the second term in brackets is greater or equal to
zero for all $\sigma_{2}$. Thus, $V_{eff}/m^{3}\geq\sigma_{2}^{2}/2$.
Finally, we observe that the $U(1)$ phase symmetry is preserved for
$\eta\leq\eta_{c}$ and is spontaneously broken by radiative corrections
for $\eta>\eta_{c}$ .

\section{Conclusions\label{sec:Sec4}}

We deform the Wess-Zumino model by implementing higher space derivatives
(i.e. Lifshitz-like operators) in the kinetic Lagrangian of it. This
is done according to the weighted renormalization condition and the
(super)symmetry requirements. In order to verify the consistency of
the model, the susy algebra is explicitly constructed by using the
Noether method in the canonical formalism. In addition, this model
is rephrased in the Lifshitz superspace (a minor and necessary modification
of the conventional one). By computing the one-loop effective potential
and analyzing their minima, we conclude that supersymmetry is preserved
at one-loop order, while the $U\left(1\right)$ phase symmetry becomes
spontaneously broken above a critical value of the renormalization
point. To study the structure of the UV divergences, we regularized
the one-loop effective potential by means of a two-dimensional cutoff
$\left(\Lambda\right)$. As expected, it is observed an improvement
of the UV behavior of the theory. Indeed, the susy-Lifshitz residual
divergence in the one-loop effective potential is logarithmic (and
not linear as in the relativistic Wess-Zumino model). This residual
divergence was removed by introducing a mass-type counterterm of the
form $A\sigma_{1}\sigma_{2}$. At this point, it is important to point
out, however, that the UV improvement depends on the exact cancellation
of the ``dangerous'' quadratic divergences between the bosonic and
fermionic contributions. It is not clear for us if this cancellation
holds at higher orders of the perturbation expansion. So, a further
study is necessary to clarify it. Finally, the construction of gauge
susy theories of the Lifshitz type is still a challenge and an open
field of research. In Appendix \ref{sec:Appendix.C}, we took a step
forward by implementing Lifshitz operators in the component formulation
of the three dimensional susy Maxwell theory. The construction of
the Lifshitz-like version of the relativistic higher-derivative SQED$_{3}$
\cite{Gallegos =000026 Baptista (2015)} is in progress and will be
reported soon.
\begin{acknowledgments}
This work was supported in part by the CNPq project No. 455278/2014-8.
The author is grateful to the referee for many valuable comments and
suggestions. Also, I would like to thank Alberto Tonero for useful
discussions during his stay in Poços de Caldas, Minas Gerais, Brazil.
\end{acknowledgments}

\appendix

\section{Lifshitz superspace in three dimensions\label{sec:Appendix.A}}

The Lifshitz superspace in its simplest way is parameterized, as in
the standard case, by three bosonic coordinates $x^{\mu}$ and two
fermionic (Grassmann) coordinates $\theta^{\alpha}$. Given that the
time and space coordinates in Lifshitz field theories are weighted
differently, we must split the Lifshitz superspace in two sectors,
one generated by the time supercharge $Q_{t\alpha}$ and the other
by the space supercharge $Q_{s\alpha}$: 
\begin{equation}
2Q_{t\alpha}=i\partial_{\alpha}+2\left(\gamma^{0}\right)_{\alpha\beta}\theta^{\beta}\partial_{0}\qquad\qquad2Q_{s\alpha}=i\partial_{\alpha}+2a\left(\gamma^{i}\right)_{\alpha\beta}\theta^{\beta}\partial_{i}.\label{eq:V.1}
\end{equation}
These supercharges in turn allow us to introduce two derivatives $D_{t\alpha}$
and $D_{s\alpha}$ defined in (\ref{eq:II.22a}) by demanding the
anticommutativity of these with the respective supercharges: $\left\{ D_{t\alpha},Q_{t\beta}\right\} =0$
and $\left\{ D_{s\alpha},Q_{s\beta}\right\} =0$. It is important
to note that these supercharges do not realize the susy algebra (\ref{eq:II.18}),
and so they are not covariant in the entire Lifshitz superspace. The
supercharge $\mathcal{Q}_{\alpha}$ that realizes (\ref{eq:II.18})
is
\begin{equation}
\mathcal{Q}_{\alpha}=Q_{t\alpha}+Q_{s\alpha}=i\partial_{\alpha}+\left(\gamma^{0}\right)_{\alpha\beta}\theta^{\beta}\partial_{0}+a\left(\gamma^{i}\right)_{\alpha\beta}\theta^{\beta}\partial_{i},\label{eq:V.2}
\end{equation}
and the covariant derivative $\mathcal{D}_{\alpha}$ with regard to
it is given by (\ref{eq:II.22b}): $\left\{ \mathcal{D}_{\alpha},\,\mathcal{Q}_{\beta}\right\} =0$.
The susy transformations (\ref{eq:II.2}) in superfield terms can
be encapsulated in the equation
\begin{equation}
\delta\Phi=i\epsilon^{\alpha}Q_{t\alpha}\Phi+i\epsilon^{\alpha}Q_{s\alpha}\Phi=i\epsilon^{\alpha}\mathcal{Q}_{\alpha}\Phi.\label{eq:V.3}
\end{equation}

In this superspace formulation, the projection technique can be implemented
in three completely equivalent ways. In fact, considering the scalar
superfield $\Phi$ in (\ref{eq:II.21}), it is easy to show that
\begin{equation}
\varphi=\left.\Phi\right|\qquad\psi_{\alpha}=2\left.D_{t,s\alpha}\Phi\right|=\left.\mathcal{D}_{\alpha}\Phi\right|\qquad F=4\left.D_{t,s}^{2}\Phi\right|=\left.\mathcal{D}^{2}\Phi\right|\label{eq:V.4}
\end{equation}
where the vertical bar $|$ means evaluation at $\theta=0$. Using
the projection technique, one gets
\begin{equation}
\int dtd^{2}xd^{2}\theta\mathscr{L}=\int dtd^{2}x\left.\left(4D_{t,s}^{2}\mathscr{L}\right)\right|=\int dtd^{2}x\left.\mathcal{D}^{2}\mathscr{L}\right|.\label{eq:V.5}
\end{equation}
The 'covariant' derivatives satisfy the following identities:
\[
\left\{ D_{t\alpha},D_{t\beta}\right\} =i\left(\gamma^{0}\right)_{\alpha\beta}\partial_{0}\qquad\qquad\left\{ D_{s\alpha},D_{s\beta}\right\} =ia\left(\gamma^{i}\right)_{\alpha\beta}\partial_{i}
\]
\[
\left\{ \mathcal{D}_{\alpha},\mathcal{D}_{\beta}\right\} =2i\left(\gamma^{0}\right)_{\alpha\beta}\partial_{0}+2ai\left(\gamma^{i}\right)_{\alpha\beta}\partial_{i}
\]
\begin{equation}
D_{t}^{2}D_{t\alpha}=-D_{t\alpha}D_{t}^{2}=\frac{i}{2}\left(\gamma^{0}\right)_{\alpha\beta}\partial_{0}D_{t}^{\beta}\qquad D_{s}^{2}D_{s\alpha}=-D_{s\alpha}D_{s}^{2}=\frac{a}{2}i\left(\gamma^{i}\right)_{\alpha\beta}\partial_{i}D_{s}^{\beta}
\end{equation}
\[
\mathcal{D}^{2}\mathcal{D}_{\alpha}=-\mathcal{D}_{\alpha}\mathcal{D}^{2}=i\left(\gamma^{0}\right)_{\alpha\beta}\partial_{0}\mathcal{D}^{\beta}+ia\left(\gamma^{i}\right)_{\alpha\beta}\partial_{i}\mathcal{D}^{\beta}
\]
\[
\left(D_{t}^{2}\right)^{2}=\frac{1}{4}\partial_{0}^{2}\qquad\left(D_{s}^{2}\right)^{2}=\frac{a^{2}}{4}\Delta\qquad\left(\mathcal{D}^{2}\right)^{2}=\partial_{0}^{2}+a^{2}\Delta,
\]
where $\partial_{0}^{2}=\partial^{0}\partial_{0}$ and $\Delta=\partial^{i}\partial_{i}$.

\section{$\mathcal{N}=1$ $\mathcal{D}=4$ Lifshitz-Wess-Zumino model\label{sec:Appendix.B}}

In this Appendix, we apply the main ideas of our three dimensional
research to implement consistently Lifshitz-like operators in the
conventional Wess-Zumino model in four spacetime dimensions. In this
case, the four dimensional spacetime manifold $\mathcal{M}_{4}$ is
split into the product $\mathbb{R}\otimes\mathcal{M}_{3}$ where $\mathbb{R}$
is the one-dimensional time manifold and $\mathcal{M}_{3}$ the three-dimensional
spatial manifold. After this separation, the Lorentz group $SO\left(1,3\right)$
reduces to the group of spatial rotations $SO\left(3\right)$.

The simplest $\mathcal{N}=1$ Lifshitz superspace in four spacetime
dimensions is parametrized by four bosonic spacetime coordinates $x^{0}=t$,
$x^{i}$ and four fermionic coordinates $\theta_{\alpha}$, $\bar{\theta}_{\dot{\alpha}}$.
According to the usual Lifshitz prescription, we assign different
weights to the time and space coordinates, namely $\left[x^{0}=t\right]_{w}=-z$
and $\left[x^{i}\right]_{w}=-1$. Hereafter we adopt the notation
of the textbook by Wess and Bagger \cite{Wess-Bagger (1992)}.

Inspired by our result (\ref{eq:II.18}), we deform the conventional
four dimensional susy algebra \cite{Wess-Bagger (1992)} by inserting
a weighted compensator $a$ in the spatial part of it:
\begin{equation}
\left\{ \mathcal{Q}_{\alpha},\,\bar{\mathcal{Q}}_{\dot{\alpha}}\right\} =2i\sigma_{\alpha\dot{\alpha}}^{0}\partial_{0}+2ai\sigma_{\alpha\dot{\alpha}}^{i}\partial_{i}.\label{eq:VII.1}
\end{equation}
It is easy to check that this deformed susy algebra is realized by
the supercharges
\begin{equation}
\mathcal{Q}_{\alpha}=\partial_{\alpha}-i\sigma_{\alpha\dot{\alpha}}^{0}\bar{\theta}^{\dot{\alpha}}\partial_{0}-ia\sigma_{\alpha\dot{\alpha}}^{i}\bar{\theta}^{\dot{\alpha}}\partial_{i}\,\,\,\,\,\,\qquad\bar{\mathcal{Q}}_{\dot{\alpha}}=-\bar{\partial}_{\dot{\alpha}}+i\theta^{\alpha}\sigma_{\alpha\dot{\alpha}}^{0}\partial_{0}+ia\theta^{\alpha}\sigma_{\alpha\dot{\alpha}}^{i}\partial_{i},\label{eq:VII.2}
\end{equation}
where $\partial_{\alpha}=\partial/\partial\theta^{\alpha}$ and $\bar{\partial}_{\dot{\alpha}}=\partial/\partial\bar{\theta}^{\dot{\alpha}}$.
Note that in four dimensions there are two types of supercharges due
to the existence of the $\gamma_{5}$ matrix (the chirality condition)
which is absent in three dimensions. 

As is customary in the superfield formulation of susy theories, we
introduce two susy covariant derivatives $\mathcal{D}_{\alpha}$ and
$\bar{\mathcal{D}}_{\dot{\alpha}}$ by demanding their anticommutativity
with the supercharges $\mathcal{Q}_{\alpha}$ and $\bar{\mathcal{Q}}_{\dot{\alpha}}$,
i. e. 
\begin{equation}
\left\{ \mathcal{D}_{\alpha},\mathcal{Q}_{\beta}\right\} =0=\left\{ \mathcal{D}_{\alpha},\bar{\mathcal{Q}}_{\dot{\beta}}\right\} \qquad\left\{ \bar{\mathcal{D}}_{\dot{\alpha}},\mathcal{Q}_{\beta}\right\} =0=\left\{ \bar{\mathcal{D}}_{\dot{\alpha}},\bar{\mathcal{Q}}_{\dot{\beta}}\right\} .\label{eq:VII.3}
\end{equation}
In doing this, one gets
\begin{equation}
\mathcal{D}_{\alpha}=\partial_{\alpha}+i\sigma_{\alpha\dot{\alpha}}^{0}\bar{\theta}^{\dot{\alpha}}\partial_{0}+ia\sigma_{\alpha\dot{\alpha}}^{i}\bar{\theta}^{\dot{\alpha}}\partial_{i},\,\,\,\,\,\,\qquad\bar{\mathcal{D}}_{\dot{\alpha}}=-\bar{\partial}_{\dot{\alpha}}-i\theta^{\alpha}\sigma_{\alpha\dot{\alpha}}^{0}\partial_{0}-ia\theta^{\alpha}\sigma_{\alpha\dot{\alpha}}^{i}\partial_{i}.\label{eq:VII.4}
\end{equation}
These susy derivatives, moreover, obey the relation $\left\{ \mathcal{D}_{\alpha},\,\bar{\mathcal{D}}_{\dot{\alpha}}\right\} =-2i\sigma_{\alpha\dot{\alpha}}^{0}\partial_{0}-2ai\sigma_{\alpha\dot{\alpha}}^{i}\partial_{i}$. 

A quick weighted analysis of (\ref{eq:VII.1}), (\ref{eq:VII.2})
and (\ref{eq:VII.4}) revels that
\begin{equation}
\left[\theta_{\alpha}\right]_{w}=\left[\bar{\theta}_{\dot{\alpha}}\right]_{w}=-z/2,\qquad\left[\mathcal{D}_{\alpha}\right]_{w}=\left[\bar{\mathcal{D}}_{\dot{\alpha}}\right]_{w}=z/2,\qquad\left[Q_{\alpha}\right]_{w}=\left[\bar{Q}_{\dot{\alpha}}\right]_{w}=z/2,\label{eq:VII.5}
\end{equation}
while the compensator $a$ weighs $\left[a\right]_{w}=z-1$. Later
on, these relations along with $\left[\partial_{i}\right]_{w}=1$
will be indispensable in the construction of the four dimensional
Lifshitz-Wess-Zumino model.

With the help of the covariant derivatives, we now define a chiral
superfield $\Phi$ by imposing the constraint
\begin{equation}
\mathcal{\bar{D}}_{\dot{\alpha}}\Phi=0.\label{eq:VII.6}
\end{equation}
Similarly, an antichiral superfield $\bar{\Phi}$ is defined by the
constraint $\mathcal{D}_{\alpha}\bar{\Phi}=0$. These constraints
are vital to build an irreducible superfield representation of the
susy algebra by eliminating the extra component fields in a generic
superfield. 

The most general solution of (\ref{eq:VII.6}) is 
\begin{eqnarray}
\Phi\left(y^{0},\,y^{i},\,\theta\right) & = & \varphi\left(y^{0},\,y^{i}\right)+\sqrt{2}\theta\psi\left(y^{0},\,y^{i}\right)+\theta^{2}F\left(y^{0},\,y^{i}\right)\nonumber \\
 & = & \varphi\left(x\right)+i\theta\sigma^{0}\bar{\theta}\partial_{0}\varphi\left(x\right)+ia\theta\sigma^{i}\bar{\theta}\partial_{i}\varphi\left(x\right)+\frac{1}{4}\theta^{2}\bar{\theta}^{2}\left(\partial_{0}^{2}+a^{2}\Delta\right)\varphi\left(x\right)\nonumber \\
 &  & +\sqrt{2}\theta\psi\left(x\right)-\frac{i}{\sqrt{2}}\theta^{2}\partial_{0}\psi\left(x\right)\sigma^{0}\bar{\theta}-\frac{ia}{\sqrt{2}}\theta^{2}\partial_{i}\psi\left(x\right)\sigma^{i}\bar{\theta}+\theta^{2}F\left(x\right),\label{eq:VII.7}
\end{eqnarray}
where the chiral coordinates $y^{0},\,y^{i}$ are given by $y^{0}=x^{0}+i\theta\sigma^{0}\bar{\theta}$
and $y^{i}=x^{i}+ia\theta\sigma^{i}\bar{\theta}$. The antichiral
superfield $\bar{\Phi}$ which obeys the constraint $\mathcal{D}_{\alpha}\bar{\Phi}=0$
is simply the complex conjugate of it, namely
\begin{equation}
\bar{\Phi}\left(\bar{y}^{0},\,\bar{y}^{i},\,\bar{\theta}\right)=\bar{\varphi}\left(\bar{y}^{0},\,\bar{y}^{i}\right)+\sqrt{2}\bar{\theta}\bar{\psi}\left(\bar{y}^{0},\,\bar{y}^{i}\right)+\bar{\theta}^{2}\bar{F}\left(\bar{y}^{0},\,\bar{y}^{i}\right),\label{eq:VII.8}
\end{equation}
where $\bar{y}^{0}=x^{0}-i\theta\sigma^{0}\bar{\theta}$ and $\bar{y}^{i}=x^{i}-ia\theta\sigma^{i}\bar{\theta}$. 

The susy transformation of the chiral superfield $\Phi$ defined in
terms of the supercharges is given by
\begin{equation}
\delta\Phi=\left(\epsilon\mathcal{Q}+\bar{\epsilon}\bar{\mathcal{Q}}\right)\Phi,\label{eq:VII.9}
\end{equation}
where $\epsilon_{\alpha}$ and $\bar{\epsilon}_{\dot{\alpha}}$ are
constant Grassmann parameters. The supersymmetric transformations
of the components fields are
\begin{eqnarray}
\delta\varphi & = & \sqrt{2}\epsilon^{\alpha}\psi_{\alpha}\nonumber \\
\delta\psi_{\alpha} & = & i\sqrt{2}\sigma_{\alpha\dot{\alpha}}^{0}\bar{\epsilon}^{\dot{\alpha}}\partial_{0}\varphi+ia\sqrt{2}\sigma_{\alpha\dot{\alpha}}^{i}\bar{\epsilon}^{\dot{\alpha}}\partial_{i}\varphi+\sqrt{2}\epsilon_{\alpha}F\label{eq:VII.10}\\
\delta F & = & i\sqrt{2}\bar{\epsilon}_{\dot{\alpha}}\bar{\sigma}^{0\dot{\alpha}\alpha}\partial_{0}\psi_{\alpha}+ia\sqrt{2}\bar{\epsilon}_{\dot{\alpha}}\bar{\sigma}^{i\dot{\alpha}\alpha}\partial_{i}\psi_{\alpha}.\nonumber 
\end{eqnarray}

Now we will build a superinvariant action of the form
\begin{equation}
S=\int d^{8}\tilde{z}\,\mathcal{K}\left(\Phi,\,\bar{\Phi}\right)+\left[\int d^{6}\tilde{z}\,\mathcal{W}\left(\Phi\right)+h.c.\right],\label{eq:VII.11}
\end{equation}
where $d^{8}\tilde{z}=dtd^{3}xd^{2}\theta d^{2}\bar{\theta}$ is the
full supermeasure and $d^{6}\tilde{z}=dtd^{3}xd^{2}\theta$ the chiral
one. In what follows, the Kähler potential $\mathcal{K}$ and the
superpotential $\mathcal{W}$, as we shall call these functions, will
be determined by imposing the weighted renormalization condition (wrc)
and the (super)symmetry requirements. Notice first of all that the
weights of these functions are $\left[\mathcal{K}\right]_{w}=3-z$
and $\left[\mathcal{W}\right]_{w}=3$, for the supermeasures in (\ref{eq:VII.11})
weigh $\left[d^{8}\tilde{z}\right]_{w}=z-3$ and $\left[d^{6}\tilde{z}\right]_{w}=-3$.

Let us focus our attention first on the Kähler $\mathcal{K}$ potential.
A typical susy operator of $\mathcal{K}$ must have the structure
$\mathcal{K}\sim\left(\mathcal{D}_{\alpha}\right)^{N_{\mathcal{D}_{\alpha}}}\left(\partial_{i}\right)^{N_{\partial_{i}}}\left(\bar{\Phi}\Phi\right)^{N_{\Phi}/2}$.
Note that to respect the spatial rotational $SO\left(3\right)$ symmetry,
$N_{\mathcal{D}_{\alpha}}$ and $N_{\partial_{i}}$ can only take
even values, i. e. $N_{\mathcal{D}_{\alpha}},\,N_{\partial_{i}}=0,\,2,\,\ldots.$
This symmetry also requires a complete contraction between the spinor/spatial
indices of the superderivatives $\mathcal{D}_{\alpha}$ and $\partial_{i}$
which might appear in $\mathcal{K}$. Considering the Kähler potential
$\mathcal{K}$ of the usual Wess-Zumino model, i.e. $\mathcal{K}=\bar{\Phi}\Phi$,
one finds the weight of the chiral $\Phi$ superfield: $\left[\Phi\right]_{w}=\left(3-z\right)/2$.
With this result at hand, we see that the Kähler $\mathcal{K}$ potential
with the above structure turns out to be renormalizable by weighted
power counting iff the condition
\begin{equation}
\frac{z}{2}N_{\mathcal{D}_{\alpha}}+N_{\partial_{i}}+N_{\Phi}\frac{\left(3-z\right)}{2}\leq3-z\label{eq:VII.12}
\end{equation}
were satisfied. From now on, we restrict our analysis to the case
$z=2$.

A careful study of the wrc (\ref{eq:VII.12}) with $z=2$ shows that
the only admissible Kähler potential $\mathcal{K}$ corresponds to
the canonical one $\mathcal{K}=\bar{\Phi}\Phi$. Consequently, one
cannot implement superderivatives, in particular, spatial $\partial_{i}$
derivatives in the Kählerian part of the superaction (\ref{eq:VII.11}).

The canonical Kählerian action $S_{\mathcal{K}}$ with $\mathcal{K}\left(\bar{\Phi},\,\Phi\right)=\bar{\Phi}\Phi$
is therefore given by
\begin{equation}
S_{\mathcal{K}}=\int d^{8}\tilde{z}\,\bar{\Phi}\Phi=\int dtd^{3}x\left\{ \bar{\varphi}\left(\partial_{0}^{2}+a^{2}\Delta\right)\varphi-i\bar{\psi}\bar{\sigma}^{0}\partial_{0}\psi-ia\bar{\psi}\bar{\sigma}^{i}\partial_{i}\psi+\bar{F}F\right\} .\label{eq:VII.13}
\end{equation}
This result can be verified by using the projection technique where
the Grassmann measure $d^{2}\theta d^{2}\bar{\theta}$ within the
spacetime integral is replaced by $d^{2}\theta d^{2}\bar{\theta}\rightarrow\bar{\mathcal{D}}^{2}\mathcal{D}^{2}/16$
and where the component fields $\varphi$, $\psi$ and $F$ are obtained
by the projections: $\varphi=\left.\Phi\right|$, $\sqrt{2}\psi_{\alpha}=\left.\mathcal{D}_{\alpha}\Phi\right|$
and $-4F=\left.\mathcal{D}^{2}\Phi\right|$\@. In addition, the following
identities are useful in deriving (\ref{eq:VII.13}):
\begin{equation}
\left[\mathcal{D}_{\alpha},\,\bar{\mathcal{D}}^{2}\right]=-4i\sigma_{\alpha\dot{\alpha}}^{0}\partial_{0}\bar{\mathcal{D}}^{\dot{\alpha}}-4ai\sigma_{\alpha\dot{\alpha}}^{i}\partial_{i}\bar{\mathcal{D}}^{\dot{\alpha}}\qquad\left[\bar{\mathcal{D}}_{\dot{\alpha}},\,\mathcal{D}^{2}\right]=4i\sigma_{\alpha\dot{\alpha}}^{0}\partial_{0}\mathcal{D}^{\alpha}+4ai\sigma_{\alpha\dot{\alpha}}^{i}\partial_{i}\mathcal{D}^{\alpha}\label{eq:VII.14a}
\end{equation}
\begin{eqnarray}
\frac{1}{8}\left[\mathcal{D}^{2},\,\bar{\mathcal{D}}^{2}\right] & = & i\sigma_{\alpha\dot{\alpha}}^{0}\partial_{0}\bar{\mathcal{D}}^{\dot{\alpha}}\mathcal{D}^{\alpha}+ai\sigma_{\alpha\dot{\alpha}}^{i}\partial_{i}\bar{\mathcal{D}}^{\dot{\alpha}}\mathcal{D}^{\alpha}+2\left(\partial_{0}^{2}+a^{2}\Delta\right)\nonumber \\
 & = & -i\sigma_{\alpha\dot{\alpha}}^{0}\partial_{0}\mathcal{D}^{\alpha}\bar{\mathcal{D}}^{\dot{\alpha}}-ai\sigma_{\alpha\dot{\alpha}}^{i}\partial_{i}\mathcal{D}^{\alpha}\bar{\mathcal{D}}^{\dot{\alpha}}-2\left(\partial_{0}^{2}+a^{2}\Delta\right).\label{eq:VII.14b}
\end{eqnarray}

Next we pass to analyse the superpotential $\mathcal{W}$. Specifically,
we seek a superpotential of the form $\mathcal{W}\sim\left(\partial_{i}\right)^{N_{\partial_{i}}}\Phi^{N_{\Phi}}$,
with all the spatial $\partial_{i}$ derivatives contracted among
themselves in order to conserve the rotational $SO\left(3\right)$
symmetry. So, $N_{\partial_{i}}=0,\,2,\,\ldots$. The wrc states that
$\mathcal{W}$ will be renormalizable by weighted power counting iff
the condition
\begin{equation}
N_{\partial_{i}}+\frac{\left(3-z\right)}{2}N_{\Phi}\leq3\label{eq:VII.15}
\end{equation}
were satisfied. Note that for $z=1$ the superpotential $\mathcal{W}$
can be at most cubic in the superfield $\Phi$ and this cannot entail
any spatial $\partial_{i}$ derivative. That is $\mathcal{W}=g\Phi+m\Phi^{2}+\lambda\Phi^{3}$.
As expected, this is really the superpotential of the conventional
Wess-Zumino model in four spacetime dimensions. 

The situation is very different for $z=2$. One can see from (\ref{eq:VII.15})
that the most general weighted renormalizable superpotential $\mathcal{W}$
is given by
\begin{equation}
\mathcal{W}\left(\Phi\right)=g\Phi+\frac{m}{2}\Phi^{2}+\frac{b}{2}\Phi\Delta\Phi+\sum_{p=1}^{4}\frac{\lambda_{p}}{p+2}\Phi^{p+2},\qquad\qquad\text{for}\,\,z=2,\label{eq:VII.16}
\end{equation}
where $\Delta=\partial^{i}\partial_{i}$ is the three-dimensional
Laplace operator. So, in this case, we can improve the UV behavior
of the theory by implementing merely two spatial $\partial_{i}$ derivatives
in the bilinear part of the usual superpotential. 

In a nutshell, we have proved by simple weighted renormalization arguments
that the unique Lifshitz-like extension with critical exponent $z=2$
of the four-dimensional Wess-Zumino model is given by
\begin{eqnarray}
S_{L-WZ} & = & \int d^{8}\tilde{z}\,\bar{\Phi}\Phi+\left[\int d^{6}\tilde{z}\left(g\Phi+\frac{m}{2}\Phi^{2}+\frac{b}{2}\Phi\Delta\Phi+\sum_{p=1}^{4}\frac{\lambda_{p}}{p+2}\Phi^{p+2}\right)+h.c.\right]\nonumber \\
 & = & \int dtd^{3}x\left\{ \bar{\varphi}\left(\partial_{0}^{2}+a^{2}\Delta\right)\varphi-i\bar{\psi}\bar{\sigma}^{0}\partial_{0}\psi-ia\bar{\psi}\bar{\sigma}^{i}\partial_{i}\psi+\bar{F}F+\left[m\left(\varphi F-\frac{1}{2}\psi^{2}\right)\right.\right.\nonumber \\
 &  & \left.\left.+b\left(\varphi\Delta F-\frac{1}{2}\psi\Delta\psi\right)+\sum_{p=1}^{4}\lambda_{p}\varphi^{p}\left(\varphi F-\frac{\left(p+1\right)}{2}\psi^{2}\right)+gF+h.c.\right]\right\} .
\end{eqnarray}
This action is of course invariant under the susy transformations
(\ref{eq:VII.10}). 

\section{$\mathcal{N}=1$ $\mathcal{D}=3$ susy Lifshitz-Maxwell theory\label{sec:Appendix.C}}

In this Appendix we construct the Lifshitz extension of the three
dimensional susy Maxwell theory in the component formalism of it.
By this example, we show explicitly the inadequacy of the current
superfield formalism of gauge theories in the implementation of higher
spatial derivatives (Lifshitz operators). 

In the three dimensional superfield formalism, the susy Maxwell theory
is described by the superaction \cite{Gates et al (1983),Gallegos =000026 Baptista (2015)}
\begin{equation}
S_{sMax}=\frac{1}{2}\int d^{5}\tilde{z}\,W^{\alpha}W_{\alpha}=\int dtd^{2}x\left\{ -\frac{1}{8}F^{\mu\nu}F_{\mu\nu}+\lambda i\cancel{\partial}\lambda\right\} ,\label{eq:VI.1}
\end{equation}
where $W_{\alpha}=\frac{1}{2}D^{\beta}D_{\alpha}A_{\beta}$ is the
superfield strength, with $D_{\alpha}=\partial_{\alpha}+i\left(\gamma^{\mu}\right)_{\alpha\beta}\theta^{\beta}\partial_{\mu}$,
and $A_{\alpha}$ denotes the spinor superfield whose field content
is:
\begin{equation}
A_{\alpha}=\chi_{\alpha}-\theta_{\alpha}B+\frac{i}{2}\theta^{\beta}\gamma_{\beta\alpha}^{\mu}V_{\mu}-\theta^{2}\left(2\lambda_{\alpha}-i\gamma_{\alpha}^{\mu\beta}\partial_{\mu}\chi_{\beta}\right).\label{eq:VI.2}
\end{equation}
The second equality in (\ref{eq:VI.1}) can be obtained by the usual
projection technique, $\int d^{2}\theta\left(\cdot\right)\rightarrow\left.D^{2}\left(\cdot\right)\right|$,
where the superfield strength $W_{\alpha}$ has the following projections:
\begin{equation}
\left.W_{\alpha}\right|=\lambda_{\alpha}\qquad\left.D_{\alpha}W_{\beta}\right|=-\frac{1}{4}F_{\mu\nu}\left(\gamma^{\mu}\gamma^{\nu}\right)_{\alpha\beta}\qquad\left.D^{2}W_{\alpha}\right|=i\left(\gamma^{\mu}\right)_{\alpha}^{\,\,\,\beta}\partial_{\mu}\lambda_{\beta},\label{eq:VI.3}
\end{equation}
with $F_{\mu\nu}=\partial_{\mu}V_{\nu}-\partial_{\nu}V_{\mu}$.

By using the identity $D^{\alpha}D_{\beta}D_{\alpha}=0$, it is easy
to verify that $W_{\alpha}$, and so the superaction (\ref{eq:VI.1}),
is invariant under the superfield gauge transformation $\delta_{g}A_{\alpha}=D_{\alpha}K$,
where $K$ labels an arbitrary scalar superfield. In component terms,
the gauge invariance of (\ref{eq:VI.1}) is established by
\begin{equation}
\delta_{g}V_{\mu}=\partial_{\mu}\xi\qquad\qquad\qquad\qquad\delta_{g}\lambda=0,\label{eq:VI.4}
\end{equation}
where $\xi$ stands for the gauge parameter. These component gauge
transformations, of course, form part of the superfield one $\delta_{g}A_{\alpha}=D_{\alpha}K$.

The implementation of Lifshitz operators directly in the conventional
superfield formulation of the susy Maxwell theory (\ref{eq:VI.1})
simply does not work. In fact, if one tries to introduce a supervertex
$\mathscr{V}_{s}$ of the form $\mathscr{V}_{s}\sim\partial_{i}^{N_{\partial_{i}}}W^{\alpha}W_{\alpha}$
, with $N_{\partial_{i}}$ extra spatial derivatives, it is found
by the wrc (noting that $\left[W_{\alpha}\right]_{w}=1$) that this
kind of vertex becomes renormalizable only for $N_{\partial_{i}}=0$,
irrespective of the value of $z$. Note that this result does not
change by modifying slightly the susy covariant derivative $D_{\alpha}$
as in (\ref{eq:II.22b}) or by rescaling the field components in (\ref{eq:VI.2}).
Note also that if $\mathscr{V}_{s}$ were renormalizable on weighted
power counting grounds it would give rise to undesirable time-space
mixing higher derivatives. 

But until what point is this superspace result conclusive to prevent
the possibility of introducing Lifshitz operators in susy gauge theories?
To answer this question, it is important to recognize and acknowledge
which the usual superfield formalim was invented to construct \textit{relativistic}
susy theories and not susy theories of Lifshitz type. We believe strongly
this superfield result is inconclusive and so the current superfield
formalism must be avoided in the construction of susy theories of
Lifshitz type.

To justify in part our assertions, we shall introduce Lifshitz operators
in the component formulation of the susy Maxwell theory (\ref{eq:VI.1})
by modifying judiciously the usual susy transformations. A simple
weighted analysis of the gauge transformation of the vector $V_{\mu}$
potential in (\ref{eq:VI.4}) reveals that the time and space components
of $V_{\mu}$ must necessarily have different weights in an eventual
theory with $z>1$. Specifically, the relation
\begin{equation}
\left[V_{0}\right]_{w}-\left[V_{i}\right]_{w}=z-1\label{eq:VI.5}
\end{equation}
must be obeyed. Hence, a Lifshitz extension of (\ref{eq:VI.1}) should
have the form
\begin{equation}
S_{sL-Max}=\int dtd^{2}x\left[-\frac{1}{4}F^{0i}F_{0i}-\frac{c^{2}}{8}F^{ij}F_{ij}+\lambda i\cancel{\partial}_{0}\lambda+c\lambda i\cancel{\partial}_{i}\lambda+\mathcal{L}_{LO}\right],\label{eq:VI.6}
\end{equation}
where $c$ stands for a weighted compensator with $\left[c\right]_{w}=z-1$
and $\mathcal{L}_{LO}$ embodies the Lifshitz operators to be implemented.
Note that the field strength $F_{\mu\nu}$ is split into two parts,
$F_{0i}$ and $F_{ij}$, since $\left[F_{0i}\right]_{w}\neq\left[F_{ij}\right]_{w}$
for $z>1$. Note also that taking $c\rightarrow1$ and $\mathcal{L}_{LO}\rightarrow0$,
one recovers the relativistic susy Maxwell theory (\ref{eq:VI.1}).

Turning off $\mathcal{L}_{LO}$, i.e. taking $\mathcal{L}_{LO}=0$,
it is not difficult to prove that (\ref{eq:VI.6}) is invariant under
the following susy transformations
\begin{eqnarray}
\delta V_{0} & = & c\epsilon i\gamma_{0}\lambda\nonumber \\
\delta V_{i} & = & \epsilon i\gamma_{i}\lambda\label{eq:VI.7}\\
\delta\lambda & = & -\frac{1}{4}F_{0i}\gamma^{0}\gamma^{i}\epsilon-\frac{c}{8}F_{ij}\gamma^{i}\gamma^{j}\epsilon.\nonumber 
\end{eqnarray}
Next, we want to implement the following Lifshitz operators
\begin{equation}
\mathcal{L}_{LO}=\alpha\,d^{2}\partial_{i}F^{ik}\partial_{j}F_{\,\,k}^{j}+d\,\lambda\Delta\lambda,\label{eq:VI.8}
\end{equation}
where $\alpha$ is a numerical factor (specifically $\alpha=-1/4$,
as will be seen below) and $d$ a coupling parameter with $\left[d\right]_{w}=z-2$.
The operators in (\ref{eq:VI.8}) are weighted renormalizable for
the upper bound of the critical exponent $z$, i. e. for $z=2$. This
may be viewed as follows. From the lower-derivative terms in (\ref{eq:VI.6}),
one finds easily that $\left[V_{0}\right]_{w}=z/2$, $\left[V_{i}\right]_{w}=1-z/2$
and $\left[\lambda_{\alpha}\right]_{w}=1$. With these results at
hand, one gets by examining each term in $\mathcal{L}_{LO}$ that
$\left[\alpha\right]_{w}=0$ and $\left[d\right]_{w}=z-2$. Then,
by demanding the polynomiality of the Lagrangian in (\ref{eq:VI.6}),
expressed in this case by $\left[V_{i}\right]_{w}\geq0$, one sees
that $z$ can take only two values: $z=1$ or $z=2$. As a result,
since the coupling constants in (\ref{eq:VI.8}) turn out weightless
for $z=2$, our statement is proved by the wrc. We would like to mention
that the first operator in (\ref{eq:VI.8}) was not chosen at random.
This indeed represents the Lifshitz version of the Lee-Wick operator,
$\partial_{\mu}F^{\mu\rho}\partial_{\nu}F_{\,\,\rho}^{\nu}$, which
characterizes the four dimensional Lee-Wick quantum electrodynamics
\cite{Lee-Wick (1970)}.

As can be explicitly shown by using the identity $\gamma^{\mu}\gamma^{\nu}\gamma^{\rho}=\gamma^{\mu\nu\rho}-\eta^{\mu\nu}\gamma^{\rho}-\eta^{\nu\rho}\gamma^{\mu}+\eta^{\mu\rho}\gamma^{\nu}$,
the action (\ref{eq:VI.6}) with $\mathcal{L}_{LO}$ defined in (\ref{eq:VI.8})
is no longer invariant under the susy transformations (\ref{eq:VI.7}).
This negative result that would discourage us from implementing $\mathcal{L}_{LO}$
was expected, however, from our Lifshitz-Wess-Zumino study. Indeed,
by eliminating totally the auxiliary field $F$ from the Lifshitz-Wess-Zumino
model (i.e. from the action (\ref{eq:II.1}) which describes it and
from the susy transformations (\ref{eq:II.2}) which it obeys), one
puts in evidence the higher derivatives of the scalar $\varphi$ field.
In particular, it is remarkable to look at how the susy transformation
of the $\psi$ field is modified in the absence of auxiliary fields
and in the presence of Lifshitz operators (taking for simplicity $m=0$
and $g=0$): $\delta\psi=-b\epsilon\Delta\varphi-i\epsilon\cancel{\partial}_{0}\varphi-ia\epsilon\cancel{\partial}_{i}\varphi$.
From this result and from the fact that there are no auxiliary fields
in the susy Maxwell formulation (\ref{eq:VI.1}), we conclude that
it is necessary to modify appropriately the susy transformations (\ref{eq:VI.7})
in order to implement $\mathcal{L}_{LO}$.

To restore the supersymmetry of the action (\ref{eq:VI.6}-\ref{eq:VI.8}),
we need to modify the susy transformations of the $V_{0}$ and $\lambda$
fields in (\ref{eq:VI.7}) to be:
\begin{align}
\delta V_{0} & =-d\epsilon\gamma_{0}\gamma_{i}\partial^{i}\lambda+c\epsilon i\gamma_{0}\lambda\nonumber \\
\delta\lambda & =-\frac{d}{4}\partial_{i}F^{ij}i\gamma_{j}\epsilon-\frac{1}{4}F_{0i}\gamma^{0}\gamma^{i}\epsilon-\frac{c}{8}F_{ij}\gamma^{i}\gamma^{j}\epsilon.\label{eq:VI.9}
\end{align}
The numerical factor $\alpha$ in (\ref{eq:VI.8}) is determined by
demanding the invariance of (\ref{eq:VI.6}), $\delta S_{sL-Max}=0$,
under the modified susy transformations (\ref{eq:VI.9}) and the unmodified
transformation $\delta V_{i}=\epsilon i\gamma_{i}\lambda$. In doing
this, one finds that $\alpha=-1/4$. 

Needless to say, that the modifications of the susy transformations
at component level necessary to introduce $\mathcal{L}_{LO}$ are
impracticable within the current superfield formalism where the field
components in (\ref{eq:VI.2}) are interconnected in such a way that
the usual susy transformations (\ref{eq:VI.7}), taking $c\rightarrow1$,
must be obeyed. It should be stressed, however, that the lower-derivative
operators in (\ref{eq:VI.6}) can be found by modifying slightly the
current superfield formalism: $\gamma^{i}\rightarrow c\gamma^{i}$
(maintaining the Clifford algebra, $\left\{ \gamma^{\mu},\gamma^{\nu}\right\} =2\eta^{\mu\nu}$,
intact) and $V_{\mu}\rightarrow c^{-1}V_{\mu}$.

The coupling of this susy Lifshitz-Maxwell theory with the Lifshitz-Wess-Zumino
model in order to define the susy Lifshitz quantum electrodynamics
is left to a forthcoming paper, for this subject lies outside of the
scope of the present investigation.

\end{document}